\documentclass[a4paper,fleqn,usenatbib]{mnras}

\usepackage[T1]{fontenc}
\usepackage{ae,aecompl}

\usepackage{graphicx}	
\usepackage{amsmath}	
\usepackage{amssymb}	

\title[Long-term monitoring of NGC 3516]{Long-term optical spectral monitoring of a changing-look AGN NGC 3516 I: Continuum and broad-line flux variability}

\author[A. I. Shapovalova et al.]
{A. I. Shapovalova$^{1}$\thanks{Deceased, 2019 January 28}, L. \v C. Popovi\'c$^{2,3}$, V. L. Afanasiev$^{1}$, 
D. Ili\'c$^{4}$, A. Kova\v cevi\'c$^{4}$, \and A. N. Burenkov$^{1}$,  V. H. Chavushyan$^{5}$, 
S. Mar\v ceta-Mandi\'c$^{4}$, O. Spiridonova$^{1}$, \and J. R. Valdes$^{5}$, N. G. Bochkarev$^{6}$,
V.  Pati{\~n}o-{\'A}lvarez$^{7,5}$, L. Carrasco$^{5}$, V. E. Zhdanova$^{1}$
\\
$^{1}$Special Astrophysical Observatory of the Russian AS,
Nizhnij Arkhyz, Karachaevo-Cherkesia 369167, Russia\\
$^{2}$Astronomical Observatory, Volgina 7, 11160 Belgrade 74, Serbia\\
$^{3}$Isaac Newton Institute of Chile, Yugoslavia Branch, Volgina 7, Belgrade, Serbia\\
$^{4}$Department of Astronomy, Faculty of Mathematics, University
of Belgrade, Studentski trg 16, 11000 Belgrade, Serbia\\
$^{5}$Instituto Nacional de Astrof\'{\i}sica, \'{O}ptica y
Electr\'onica, Apartado Postal 51-216, 72000 Puebla, Puebla, M\'exico\\
$^{6}$Sternberg Astronomical Institute of Lomonosov Moscow State University, 
119992, Moscow, Russia\\
$^{7}$Max-Planck-Institut f\"ur Radioastronomie, Auf dem Hugel 69, D-53121 Bonn, Germany
}

\date{Accepted 2019 February 22. Received 2019 February 21; in original form 2018 November 29.}

\pubyear{2018}

\begin{document}
\label{firstpage}
\pagerange{\pageref{firstpage}--\pageref{lastpage}}
\maketitle

\begin{abstract}
Here we present the long-term optical spectral monitoring of a changing-look active galactic nuclei (AGN) NGC 3516 that covers 22 years (from 1996 to 2018). We explore a variability in the broad lines and continuum, finding that the continuum  is changing  by more than a factor of 2, while the broad lines are varying  by more than a factor of 10. The minimum of activity is observed in 2014, when the broad lines almost disappeared. We confirm that NGC 3516 is a changing-look AGN, and the absorption seen in the UV and X-ray may indicate that there is an obscuring region which is responsible for this.

The line profiles are also changing. The mean profiles of the broad H$\alpha$ and H$\beta$ lines show shoulder-like structure in the wings, and enhanced peak, that may indicate a complex BLR. The rms-profiles of both lines seem to have the same shape and width of around 4200 km s$^{-1}$, indicating practically the same kinematics in the H$\alpha$ and H$\beta$ emitting regions.

Measured time-lags between the continuum and H$\alpha$ and H$\beta$ broad-line variability are  $\sim$ 15 and 17 days, respectively, that in combination with the broad lines width allows us to  estimate the NGC 3516 central black hole mass. We  find that the black hole mass is 
 (4.73$\pm$1.40)$\times 10^7M_\odot$ which is in agreement with previous estimates.

\end{abstract}




\section{Introduction}

The nature of the "central engine" of active galactic nuclei (AGN) is still an open question. However,  
 usually it is assumed that the nuclear activity is caused by the accretion of matter on a super-massive black hole
 \cite[SMBH, see][]{re84,be85}. 
 The radiation from the accretion disc is ionizing the surrounding gas, that forms a so called broad  line region (BLR), which is located very close to the central  SMBH (r$<$0.1 pc). The BLR is very compact (several 10s to several 100s light days), i.e. dimensions of the BLR correspond to approximately 10$^{-4}$ arcsec in  the nearest AGN, which  is a great challenge to resolve with the current largest telescopes, e.g. for now only the BLR of a quasar 3C 273 is resolved using the GRAVITY interferometer \citep[][]{st18}. Therefore,  spectroscopy  and/or spectro-polarimetry can give  valuable information about the  BLR structure  in a larger number of AGN  \citep[see][]{af18}.
 Over the past 40 years, the variability of broad emission lines and  continuum in the majority of AGN has been detected.
Already from the pioneering works in the seventies \citep[see][]{ch73,bo78}
it became clear that the intensities of the broad emission 
lines in AGN change with a time delay of 1-3 weeks with respect to  the continuum change. 
The time delay depends on the time of passage of light through the
BLR and on the BLR geometry and kinematics  \citep[][]{bo82,bl82}
This can be used for investigations of the  BLR structure and dimension, 
i.e. by finding correlations between  changes in the continuum and broad-line fluxes 
it is possible to "map" the  BLR structure. This method is a so called reverberation method \citep[see][and references therein]{pe93}. 
Additionally, one can explore the BLR evolution by studying the variability of the broad emission  lines on a long time scale, i.e.  study 
the changes in the BLR physics, kinematics and geometry as a function of time. Finally, the BLR is supposed to be near to  the
SMBH in AGN and may hold basic information about the formation and fueling of AGN, and especially can give information about the 
central black hole mass in the case of the BLR virialization.

Here we  study  the variability of the broad emission  lines and  continuum of active galaxy NGC 3516  on a long
time scale (1996--2018).  NGC 3516 was one of the first observed active galaxies
 that showed a variable flux in the broad lines \citep[][]{an68}
The galaxy is a close (z$\sim$0.009),  bright (V$\sim$12.5 magnitude, but also  variable) object with  the morphological  type SBO. The optical spectrum 
of the NGC 3516 nucleus was studied repeatedly \citep[see][and reference therein]{co73,ad75,bo77,os77,cr86,wa93}.
 Strong variations in the intensity of the broad lines  and optical Fe II lines were reported in several papers
\citep[][]{so68,an71,co73,bo77,co88,bo90}. 
In NGC 3516 the contribution of the absorption spectrum of the galactic nucleus stellar component is very  significant \citep[][]{cr85}, and in addition, NGC 3516 shows a strong intrinsic UV absorption, which is  blueshifted  \citep[][]{go99}.
The absorption lines width and ionization state are consistent with one expected in the narrow line region (NLR), i.e. 
it seems that the origin of the  UV absorption in NGC 3516 is in the NLR \citep[see][]{go99}. 

This significant contribution of the stellar population to the AGN continuum, estimated from the aperture of the size of 1.0$\arcsec\times 4.0\arcsec$, is  $\sim$70\% to the continuum flux in the H$\beta$ wavelength region \citep[][]{bo90}.
This makes NGC 3516 potentially a very interesting AGN, since there is a huge 
contribution of the circum-nuclear component, and  in addition,  in the center resides a low luminosity AGN that emits the broad Hydrogen 
emission which  is strongly  time-variable.  The observed H$\alpha$/H$\beta$  \citep[$\sim$5, see e.g.][]{de16} is larger than the theoretical Case B value, which is expected from the pure photoionization model \citep[][]{of06}. However, a special case of photoionizaton model could explain the observed Balmer lines ratio \citep[][]{de16}, as well as some alternative approaches \citep[][]{po02}. In addition, the effects of collisional excitation and dust extinction could be the reason of such large deviation of the Balmer decrement.

In 1990 the first tight spectral and photometric optical monitoring during 5 months was performed as part of the LAG (Lovers of Active Galaxies) collaboration \citep[][]{wa93,wa94,on03}.
A large amplitude of variability of broad lines and continuum, variable asymmetric line profiles, and a variable dip in the blue wing of H$\beta$ were detected,  and  time-lags were also estimated \citep[H$\alpha$-14 and H$\beta$-7 days][]{wa94}.  In 2007 a high sampling rate, 
6-month optical reverberation mapping campaign of NGC 3516,  was undertaken at MDM Observatory with the support of observations at several telescopes \citep[][]{de10}.
They showed that the H$\beta$ emission region within the BLR  of NGC 3516  has complex kinematics (clearly see evidence 
for outflowing, infalling, and virialized BLR) and  reported an updated  time-delay of the broad H$\beta$ line (11.7 days). Additionally, the line shape investigation given by \cite{po02} indicated a presence of a disc-like BLR which emits mostly in the line wings, and 
 another  BLR component  (so called intermediate BLR - IBLR) which emits narrower lines and contributes to the line core. Recently, \cite{de18} found that the time delay between the continuum and H$\beta$ is $\sim$4-8 days, that in combination with the measured  root-mean-square (rms) profile of H$\beta$ width (around 2440 km s$^{-1}$) gives the central black hole mass of $\log(M/M_\odot)=7.63$.
Finally, NGC 3516 was under the simultaneous monitoring in the X-ray and optical B-band in 2013--2014 \citep[][]{no16},
when the object was detected in its faint phase.

In this paper (Paper I), we present the results  of the  long-term photometric (B,V,R) and  spectral 
(in the H$\alpha$ and H$\beta$ wavelength band) monitoring of NGC 3516 during the period between 1996 and 2018, and discuss the broad line and continuum flux variability.  In Paper II we are going to investigate the changes in the BLR, i.e. in the shape of broad lines. 
The paper is organized as follows: in Section 2 we report on our observations and describe the data reduction; in 
Section 3 we describe the performed data analysis, and in Section 4 we discuss our results; finally in Section 5 we outline our conclusions.

\begin{figure}
\centering
\includegraphics[width=\columnwidth]{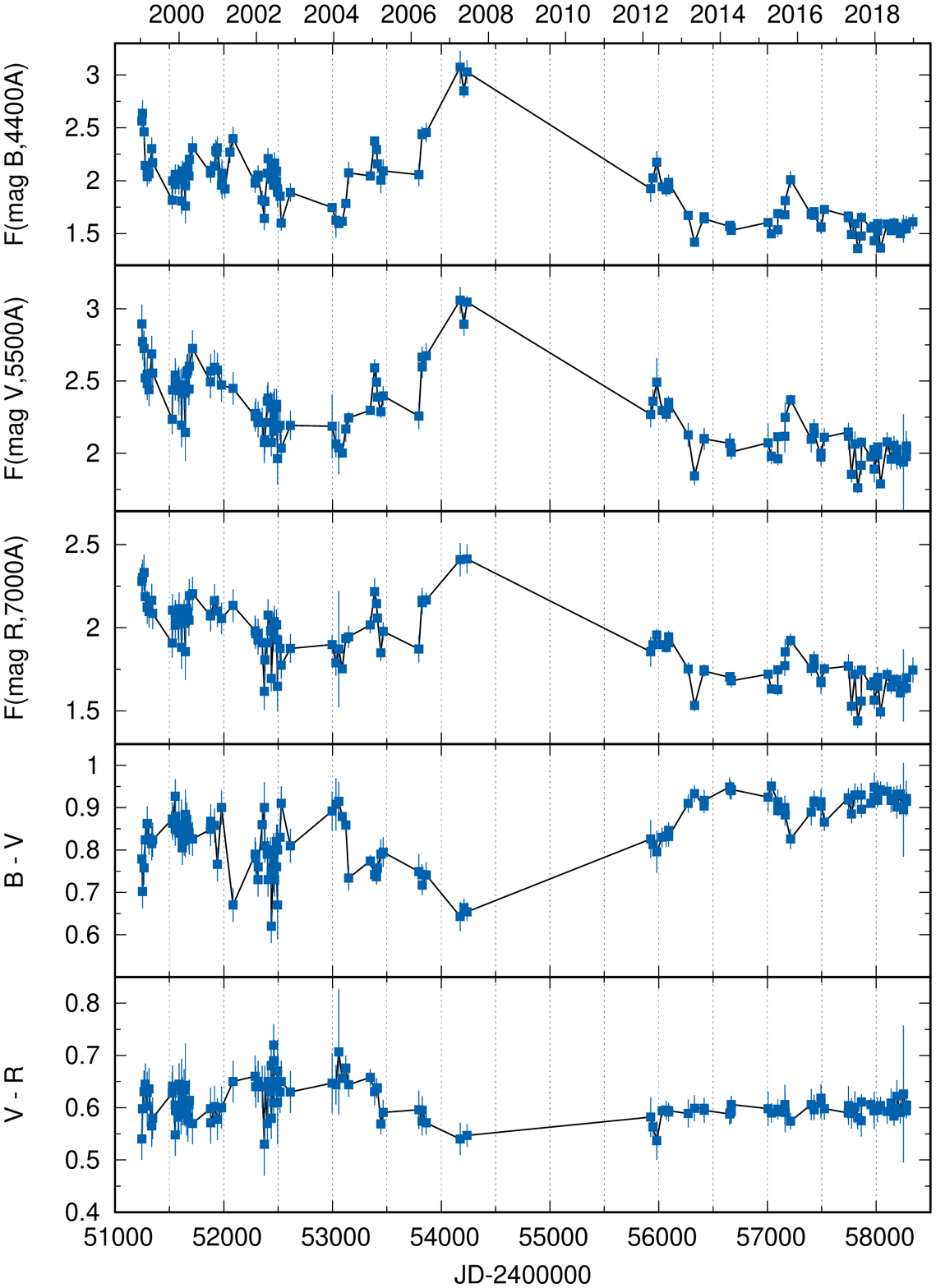}
\caption{Photometric light curves in B, V, R filters, transformed into corresponding fluxes using the equations from \citet{di98}. The fluxes are given in units of
10$^{-15}$ erg cm$^{-2}$s$^{-1}$\AA$^{-1}$. Bottom two panels give the light curves of
color indexes B-V and V-R, in stellar magnitudes.} \label{fig1}
\end{figure}

\begin{table*}
\centering
\caption{ A sample of measured photometric magnitudes of NGC 3516. Columns are: (1): Number, (2): Modified Julian date (MJD), (3): Mean seeing in arcsec, and (4)-(6): BRV magnitudes and corresponding errors. The full table is available online as Supporting information.}
\label{tab1}
\begin{tabular}{cccccc}
\hline
N & MJD & Seeing & m$_{\rm B}$ $\pm \sigma$ & m$_{\rm V}$ $\pm \sigma$ & m$_{\rm R}$ $\pm \sigma$  \\
  & 2400000+ & [arcsec] &  &    &   \\
1 & 2 & 3 & 4 & 5 & 6  \\
\hline
1	& 52996.63 & 1.3 & 13.945$\pm$0.002 & 13.053$\pm$0.044 & 12.406$\pm$0.013 \\
2	& 53031.55 & 2.0 & 14.023$\pm$0.043 & 13.115$\pm$0.019 & 12.471$\pm$0.019 \\
3	& 53058.48 & 2.5 & 14.044$\pm$0.007 & 13.129$\pm$0.039 & 12.422$\pm$0.081 \\
4	& 53088.44 & 3.0 & 14.028$\pm$0.006 & 13.149$\pm$0.004 & 12.493$\pm$0.000 \\
5	& 53122.23 & 3.0 & 13.921$\pm$0.004 & 13.062$\pm$0.018 & 12.386$\pm$0.006 \\
6	& 53149.33 & 1.5 & 13.758$\pm$0.022 & 13.024$\pm$0.008 & 12.380$\pm$0.015 \\
7	& 53347.57 & 3.0 & 13.773$\pm$0.009 & 12.999$\pm$0.005 & 12.341$\pm$0.011 \\
8	& 53386.55 & 1.2 & 13.611$\pm$0.000 & 12.869$\pm$0.010 & 12.238$\pm$0.016 \\
9	& 53405.49 & 2.2 & 13.648$\pm$0.007 & 12.911$\pm$0.012 & 12.274$\pm$0.007 \\
10	& 53413.47 & 2.2 & 13.714$\pm$0.018 & 12.957$\pm$0.004 & 12.319$\pm$0.005 \\
\hline
\end{tabular}
\end{table*}

\section{Observations and data reduction}
\label{sec:obs}

Details about the observations, calibration and unification  of the spectral data, and measurements of the spectral fluxes are reported in our previous works \citep[see][and references therein]{sh01,sh04,sh08,sh10,sh12,sh13, sh16,sh17}, and will not be repeated here. However, we give some basic information about photometric and spectral observations of NGC 3516 and data reduction.

\begin{table*}
\centering
\caption{Details of the spectroscopic observations. Columns are: (1): Observatory, (2): Code, (3): Telescope aperture and type of spectrograph. (4): Projected spectrograph
entrance apertures (slit width$\times$slit length in arcsec), and (5): Focus of the telescope.}
\label{tab2}
\begin{tabular}{lcccc}
\hline
Observatory & Code & Tel.aperture + equipment & Aperture [arcsec] & Focus \\
1 & 2 & 3 & 4 & 5\\
\hline
SAO (Russia)  & L(N) & 6 m + Long slit  &  2.0$\times$6.0     &      Nasmith    \\
SAO (Russia)  & L(U) & 6 m + UAGS       &  2.0$\times$6.0     &      Prime      \\
GHO (M\'exico)& GHO  & 2.1 m + B\&C      &  2.5$\times$6.0     &      Cassegrain \\  
SAO (Russia)  & Z1   & 1 m + UAGS       &  4.0$\times$19.8    &      Cassegrain \\
SAO (Russia)  & Z2K  & 1 m + UAGS       &  4.0$\times$9.45    &      Cassegrain \\
\hline
\end{tabular}
\end{table*}

\begin{table*}
\centering
\caption{Spectroscopic observations log. Columns are: (1): Number, (2): UT date, (3): Modified Julian date (MJD), (4): Code, given in Table~\ref{tab2}, (5): Projected spectrograph entrance apertures, (6): Wavelength range covered, and (7): Mean seeing in arcsec. The full table is available online as Supporting information.}
\label{tab3}
\begin{tabular}{ccccccc} 
\hline
N & UT-date & MJD & Code & Aperture &Sp.range & Seeing  \\
  &    &2400000+ &  & [arcsec] & [\AA]   &  [arcsec] \\
1 & 2 & 3 & 4 & 5 & 6 & 7\\
\hline
  1 & 14.01.1996 & 50096.63 & Z1   & 4.0$\times$19.8& 3738-6901 &  4.0   \\
  2 & 20.01.1996 & 50103.47 & L(N) & 4.0$\times$19.8& 5100-7300 &  -     \\
  3 & 19.03.1996 & 50162.31 & L(N) & 2.0$\times$6.0 & 3702-5595 &  3.0   \\
  4 & 20.03.1996 & 50163.35 & L(N) & 2.0$\times$6.0 & 3702-5595 &  4.5   \\
  5 & 05.10.1997 & 50726.62 & L(N) & 2.0$\times$6.0 & 3845-6288 &  4.0   \\
  6 & 07.10.1997 & 50728.64 & L(N) & 2.0$\times$6.0 & 3845-6289 &  4.0   \\
  7 & 20.01.1998 & 50834.34 & L(N) & 2.0$\times$6.0 & 3838-6149 &  2.5   \\
  8 & 28.01.1998 & 50842.44 & L(U) & 2.0$\times$6.0 & 4540-5348 &  2.8   \\
  9 & 22.02.1998 & 50867.31 & L(N) & 2.0$\times$6.0 & 3837-6149 &  2.0   \\
 10 & 07.05.1998 & 50940.53 & L(N) & 2.0$\times$6.0 & 3738-6149 &  3.0   \\
\hline
\end{tabular}
\end{table*}

\begin{table*}
\centering
\caption{Flux scale factors $\varphi$ and extended source correction G(g) 
[in units of 10$^{-15} \rm erg \ cm^{-2} s^{-1}$\AA$^{-1}$] for the optical 
spectra in the case of different telescopes. GHO(m) sample contains spectra with 
spectral resolution of 15 \AA.}
\label{tab4}
\begin{tabular}{lcccc}
\hline
Sample & Years & Aperture& Scale factor& Extended source  correction \\
    &    &    (arcsec) & ($\varphi\pm\sigma$) & G(g)  \\
\hline
  GHO    & 1999-2007 &  2.5$\times$6.0  &  1.000           &   0.000         \\
  GHO(m) & 1999-2007 &  2.5$\times$6.0  &  1.020$\pm$0.085 &   0.000          \\
  L(U,N) & 1999-2010 &  2.0$\times$6.0  &  1.230$\pm$0.049 &   1.42$\pm$1.18  \\
  Z1K    & 1999-2004 &  4.0$\times$19.8 &  1.350$\pm$0.110 &   6.58$\pm$0.73  \\
  Z2K    & 2003-2017 &  4.0$\times$9.45 &  1.319$\pm$0.072 &   5.92$\pm$2.45  \\
\hline
\end{tabular}
\end{table*}

\subsection{Photometric observations}
\label{sec:phot}

The photometry in the BVR filters of NGC 3516 was performed at the Special Astrophysical Observatory of the 
Russian Academy of Science (SAO RAS) during  the 1999 -- 2017 period (139 nights)  with CCD-photometers of 1-m and 60-cm Zeiss telescopes.
The photometric system is similar to those of Johnson in B and V, and of Cousins in R spectral band \citep[][]{co76}.
The software developed at SAO RAS by \cite{vl93}
was used for the data reduction.  Photometric standard stars from \cite{pe71},
in 1998--2003, and from \cite{do05},
in 2004--2017, were used.  In Table \ref{tab1} the photometric BVR-magnitude data for the aperture of 10\arcsec are presented. In Figure \ref{fig1} we plot the light curves in the BVR bands and (B-V), (V-R) color indexes.
For the light curves (Fig. \ref{fig1}), the magnitudes [m(B), m(V),m(R)] were transformed into 
fluxes F(B), F(V) and F(R) in units of 10$^{-15}$ erg cm$^{-2}$ s$^{-1}$ \AA$^{-1}$, using the equations from \cite{di98}.

\begin{figure*}
\centering
\includegraphics[width=10cm, angle=90]{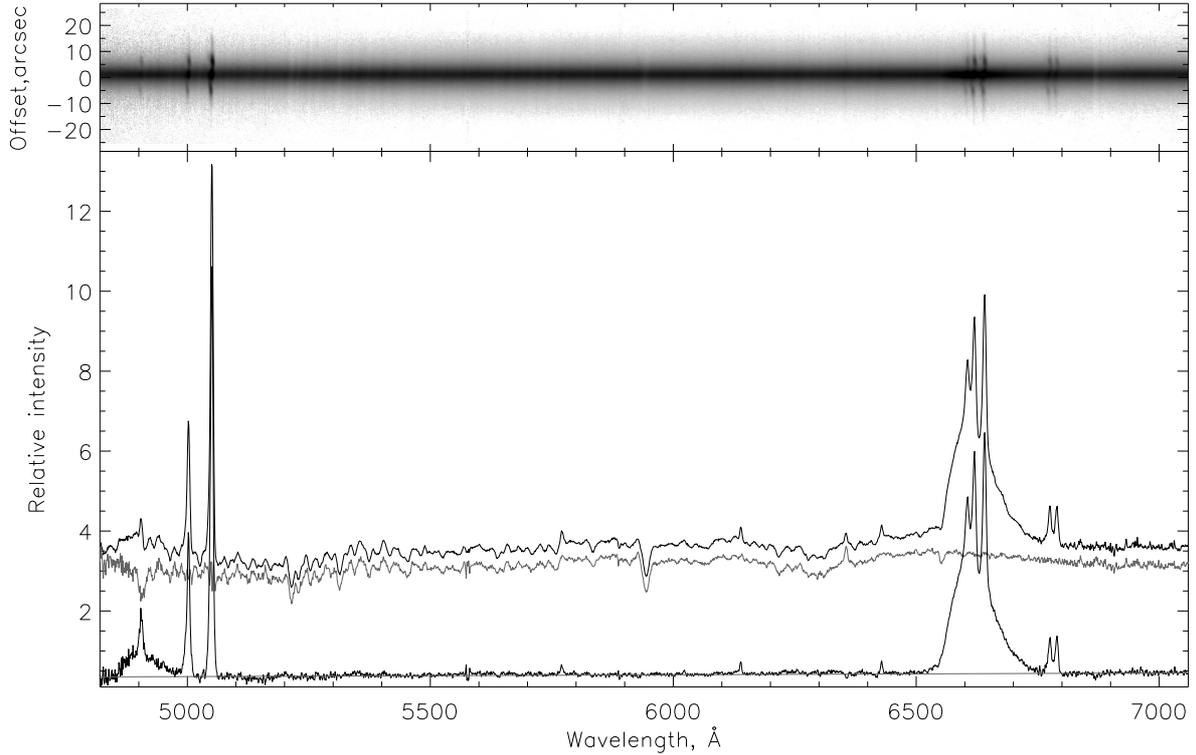}
\caption{The image of the observed spectrum of NGC 3516 in 2017 (top panel), and from top to bottom (bottom panel): the extracted composite (aperture of 2\arcsec$\times$4\arcsec) , host galaxy, and  pure AGN spectrum, respectively.} \label{sp2017}
\end{figure*}

\subsection{Spectral observations}
\label{sec:spec}

Spectral monitoring of the galaxy NGC 3516 was carried out  in 1996--2007 and 2014--2018
during $\sim$160  observing nights.  Spectra  were taken with the 6 m and 1 m telescopes
of the SAO RAS, Russia (1996--2018), and with the 2.1 m telescope of the Instituto Nacional de Astrof\'{\i}sica, \'{O}ptica y
Electr\'onica (INAOE) at the "Guillermo Haro
Observatory" (GHO) at Cananea, Sonora, Mexico (1998--2007). They were obtained with
long-slit spectrographs equipped with CCDs. The typical covered wavelength interval  was
from $\sim$3700 \AA\  to 7700 \AA,  the spectral resolution was between $\sim$(8--10) \AA \, or (12--15) \AA, and
the signal-to-noise (S/N) ratio was $\sim$40--50 in the continuum near the H$\beta$ and H$\alpha$ lines.
Spectrophotometric standard stars were observed every night. Table \ref{tab2} provides  a short information on the source of spectroscopic observations. The log of spectroscopic observations is given in Table \ref{tab3}. The spectrophotometric data reduction was carried out using either the software developed at SAO RAS \citep[][]{vl93} or the
IRAF package for the spectra obtained at GHO, and it included bias and flat-field corrections, cosmic ray removal, 2D wavelength linearization, sky spectrum subtraction, addition of the spectra for every night, and relative flux calibration based on  spectrophotometric standard star observations. In the analysis, about 10\% of the spectra were discarded for several different reasons (e.g. high noise level,
badly corrected spectral sensitivity, poor spectral resolution $>$15 \AA, etc.). Thus, our final data set consists of 
123 blue (covering H$\beta$)  and 89 red (covering H$\alpha$) spectra, taken during  146 nights, which we use in further analysis.

Additionally, we observed NGC 3516 with the 6 m telescope with the SCORPIO-2 spectrograph on February 1, 2017 in the spectral range from  4820 \AA\ to 7060 \AA. The observations were done in spectro-polarimetic mode with the grating VPHG1800@590 giving a dispersion of 
0.5 \AA\ per px with a spectral resolution of 4.5 \AA . The slit width was 2\arcsec, 
and the height was 57\arcsec. The exposure time was 3600 s and seeing was 2.3\arcsec. The 
observed spectrum is shown in Figure \ref{sp2017}  (top panel), from which, due to its high-quality we 
could extract the composite spectra (aperture size of 2\arcsec$\times$4\arcsec, top spectrum, bottom panel), the 
spectrum of the host galaxy (middle spectrum, bottom panel) and the 
pure AGN spectrum (bottom spectrum, bottom panel). To subtract the host galaxy spectrum from the composite one, 
we extracted the offset spectra from two regions  in the range from  -3\arcsec to -18\arcsec below and 
from +3\arcsec' to +18\arcsec above the center.  The averaged values (middle spectrum, bottom panel) 
is subtracted from the composite spectrum, and we 
obtained the spectrum of the  pure AGN (bottom spectrum, bottom panel), which 
is showing the presence of weak broad components in the H$\alpha$ and 
H$\beta$ lines. As it can be seen from the Figure \ref{sp2017}  the narrow 
and broad lines are present.

\begin{figure}
\centering
\includegraphics[width=\columnwidth]{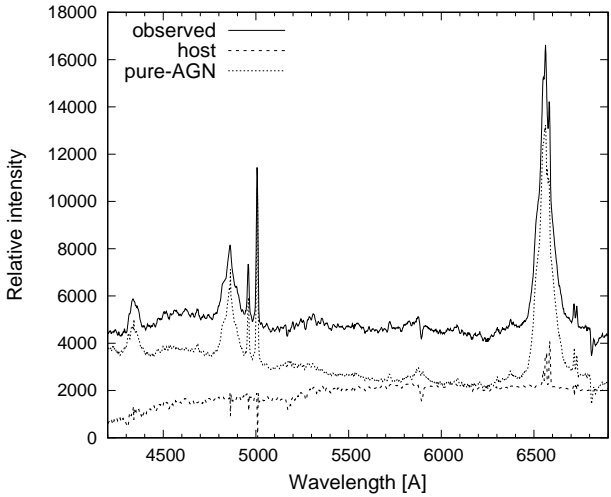}
\caption{The PCA decomposition of the year-average spectrum in 1997.} \label{pca}
\end{figure}

 The above procedure for the host-galaxy subtraction, could be done only in the case of the latest high-quality spectrum. In order to test if there is a "hidden" broad-line component in the H$\alpha$ and H$\beta$ line profiles  in  all spectra from our campaign, we estimated the host-galaxy contribution using the spectral principal component analysis (PCA), a statistical method which is described in \citet{fr92,vb06}. We applied the PCA   to the year-average spectra, obtained from those spectra covering the total wavelength range.
\citet{vb06} introduced the application of this
statistical method for spectral decomposition of a composite spectrum into a pure-host and pure-AGN part. The PCA uses eigenspectra of AGN and galaxies, whose linear combination can reproduce the observed spectrum \citep[see][etc.]{fr92,co95,yi04a,yi04b}.
An example of the PCA decomposition of the year-average observed spectrum (from 1997) to the host-galaxy  and pure-AGN spectrum is shown in Figure \ref{pca}. The obtained host-galaxy spectra were subtracted from the observed year-average spectra in order to obtain the pure AGN component.

\begin{figure}
\centering
\includegraphics[width=\columnwidth]{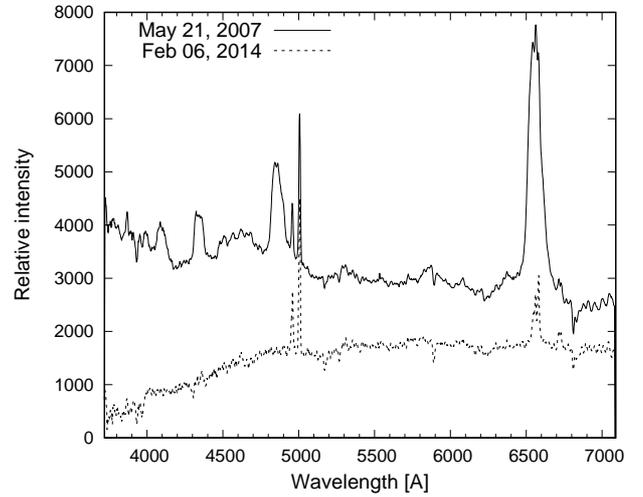}
\caption{Observed spectra in the minimum and maximum of activity during the monitored period (epoch of
observations denoted in the upper left corner).}
\label{fig-mm}
\end{figure}

\subsection{Absolute calibration (scaling) of the spectra}
\label{sec:cal}

Usually, for the absolute calibration of the spectra of AGN, fluxes in the narrow emission lines are used because it is assumed 
that they are not variable at intervals of tens of years \citep[]{pe93}.  All blue spectra of NGC 3516 were thus scaled to the constant 
flux of F([O III] $\lambda$(5007+4959) = 4.47$\times$10$^{-13}$ erg cm$^{-2}$ s$^{-1}$.  This value is obtained using the data of \cite{de10} for F([O III] $\lambda$5007)= 3.35$\times$10$^{-13}$ erg cm$^{-2}$ s$^{-1}$ and the flux ratio of F([O III] $\lambda$5007)/F([O III] $\lambda$4959) = 3 \citep[see][]{di07}.

The scaling method of the blue spectra \citep[see][]{sh04}
allows us to obtain a homogeneous set of spectra with the same wavelength calibration and the same [O III] $\lambda$4959+5007 fluxes. 
The spectra obtained using the SAO 1-m telescope with the resolution of  $\sim$8-10 \AA\  (UAGS+CCD2K, Table \ref{tab2})  and  spectra  of  2.1m telescope  with the resolution $\sim$12-15 \AA\ (Boller\&Chivens  spectrograph + a grism of 150 l/mm)  are covering both the H$\alpha$ and H$\beta$ spectral bands. These spectra were scaled using the [O III] $\lambda$4959+5007  lines, and consequently, the red spectral band was  automatically scaled to the flux of these lines. 

The blue spectra of NGC 3516 in the wavelength region of  $\sim$(3700--5800) \AA \, and with the spectral resolution of $\sim$8 \AA, taken with the 6 m and 1 m telescopes of SAO RAS and with the 2.1 m telescope of GHO\footnote{Code L(N), L(U), Z1 and GHO from Tables 2 and 3} were also scaled using the flux of the [O III] $\lambda$4959+5007  lines. The red spectra observed at the same night (or next night) in the wavelength region (5600--7600) \AA \, were first scaled to the fluxes of [S II] $\lambda$6717+6731 lines, and then the scaling was corrected using the overlapping continuum with the corresponding blue spectrum which was scaled to [O III] $\lambda$4959+5007.  The [S II] $\lambda$6717+6731 total flux was determined from the scaled spectra covering the entire wavelength range. However, the accuracy of the scaling of the red region depends both on the accuracy of the determination of the  [S II] lines flux and on the slope of the continuum. In the spectra of NGC 3516, the fluxes in the [S II] $\lambda\lambda$6717,6731 lines are almost an order of magnitude  smaller then  the fluxes in the [O III] $\lambda\lambda$4959,5007 lines. Therefore, when scaling to the [S II] $\lambda$6717+6732 flux, the scaling accuracy varied within 2-10\%, depending on the quality of the spectrum.  To improve the accuracy of the scaling, we used overlapping sections of the continuum of the blue and red spectra recorded on the same or next night. However, in this case the accuracy of the scaling procedure depends strongly on the determination of the continuum slope in the blue and red spectral band, i.e. one has to carefully account the spectral sensitivity of the equipment. This has been performed by using the comparison stars. In poor photometric conditions (clouds, mist, etc.) the reduction can give a wrong spectral slope (fall or rise) and, consequently, the errors in the scaling procedure for the H$\alpha$ wavelength band can be larger. Usually (as a rule), the fluxes in the H$\alpha$ line and red continuum  determined from the spectra scaled to the [S II] $\lambda$6717+6731 flux or using the overlapping sections of the continuum   show little difference (less than 5\%), but in several red spectra ($\sim$6\%) the fluxes differ up to 10\%. In the latter case, we used the flux from the average spectrum. Similarly, in the case of spectra that cover the whole wavelength range, which were scaled to [O III] $\lambda$4959+5007,  for better precision, we also scaled the spectra using the flux in [S II] $\lambda$6717+6731. Then we compared the fluxes in the H$\alpha$ line and the red continuum obtained from  two differently scaled spectra, and if there were differences of more than 5\%, the average flux was used. As it was mentioned at the beginning of Section 
2, more details on the scaling can be  found in our previously published papers \citep[see e.g.][]{sh01,sh04,sh08,sh10,sh12,sh13,sh16,sh17}.

\begin{figure*}
\centering
\includegraphics[width=16cm]{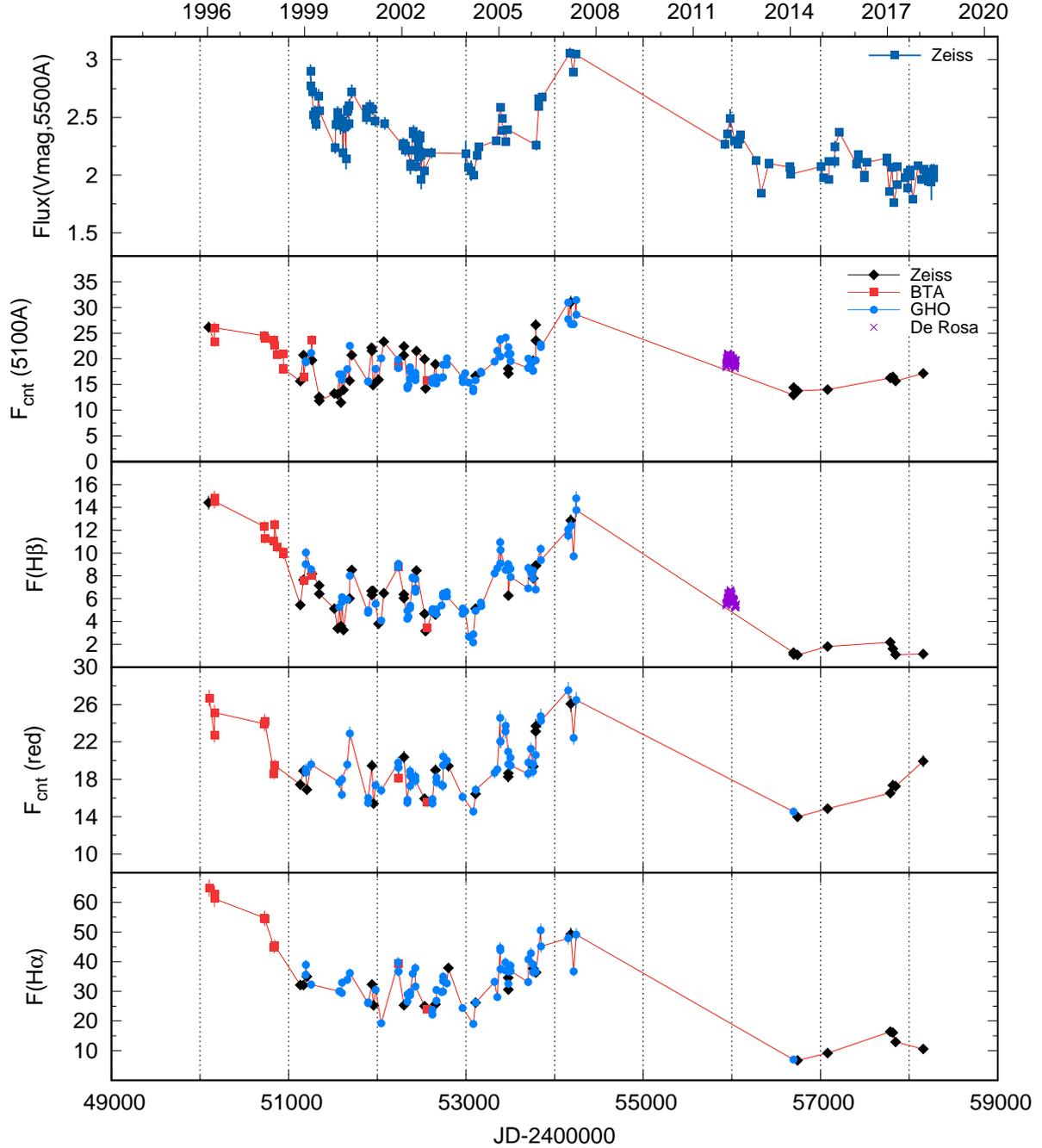}
\caption{Light curves for the spectral lines and continuum fluxes, compared to 
the photometry flux in the V filter, F(V,$\lambda$5500\AA) shown in top panel. Observations with different 
telescopes are denoted with different symbols given in the second panel from the top. Also, observations reported by \citet{de18} are included.
The continuum flux is in units of 
$10^{-15} \rm erg \, cm^{-2} s^{-1}$\AA$^{-1}$ and the line flux in units of $10^{-13} \rm erg \, cm^{-2} s^{-1}$.} \label{fig2}
\end{figure*}

\begin{figure*}
\centering
\includegraphics[width=\columnwidth]{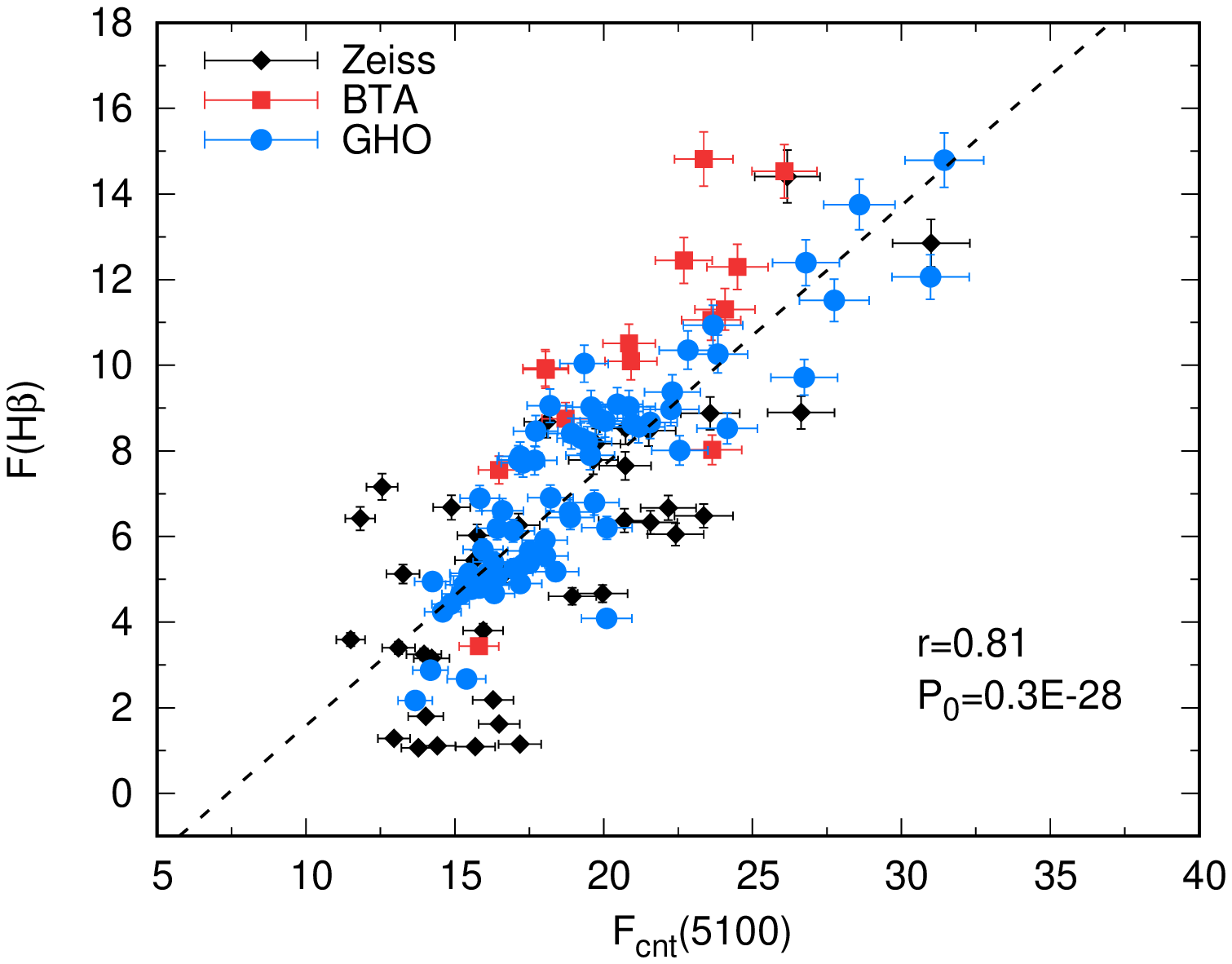}
\includegraphics[width=\columnwidth]{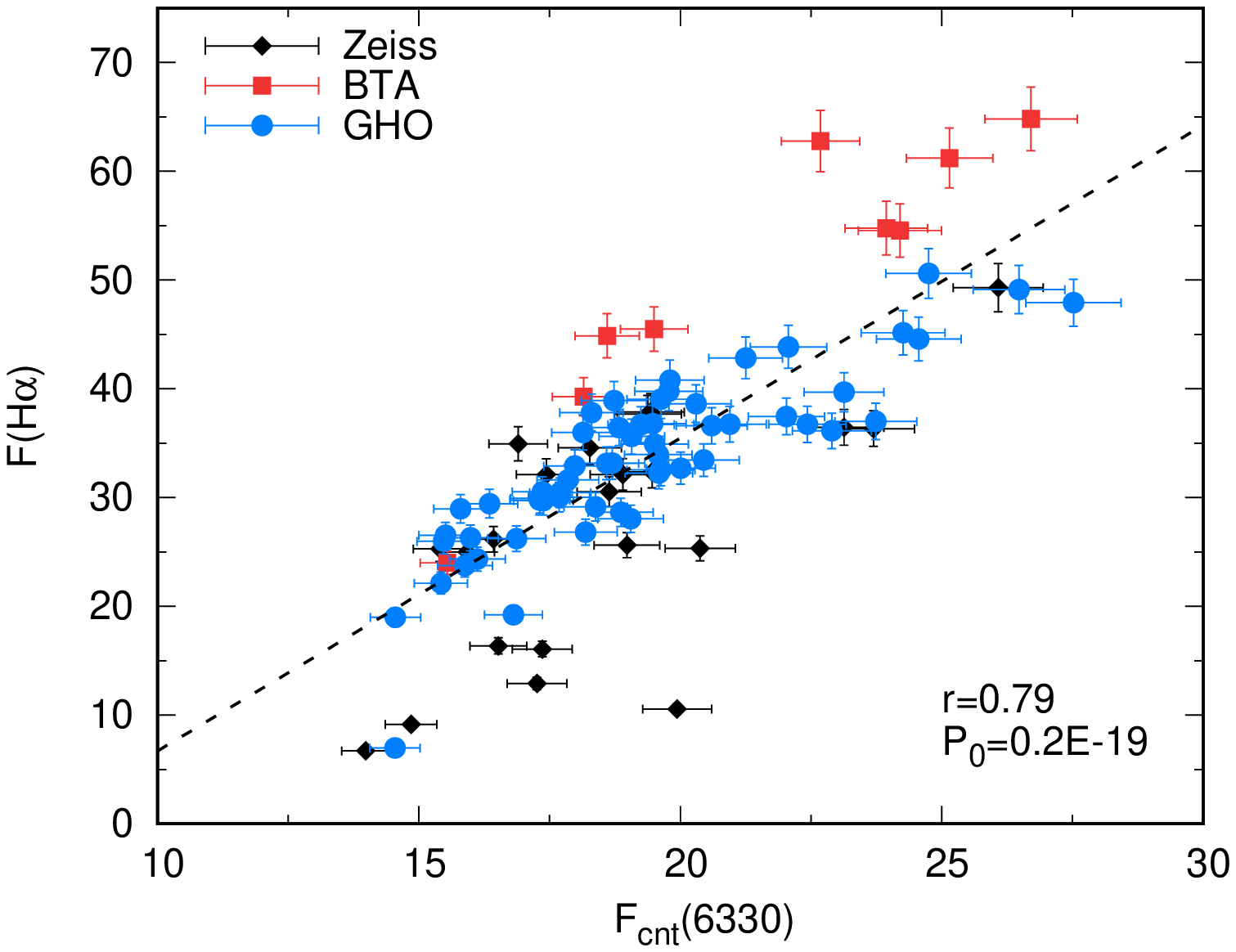}
\caption{Continuum vs. line flux for H$\beta$ and H$\alpha$. Symbols and units are the same as in Figure \ref{fig2}.
The correlation coefficients $r$ and the corresponding null-hypotesis $P_0$ values are also given.} \label{fig3}
\end{figure*}

\begin{figure*}
\centering
\includegraphics[width=\columnwidth]{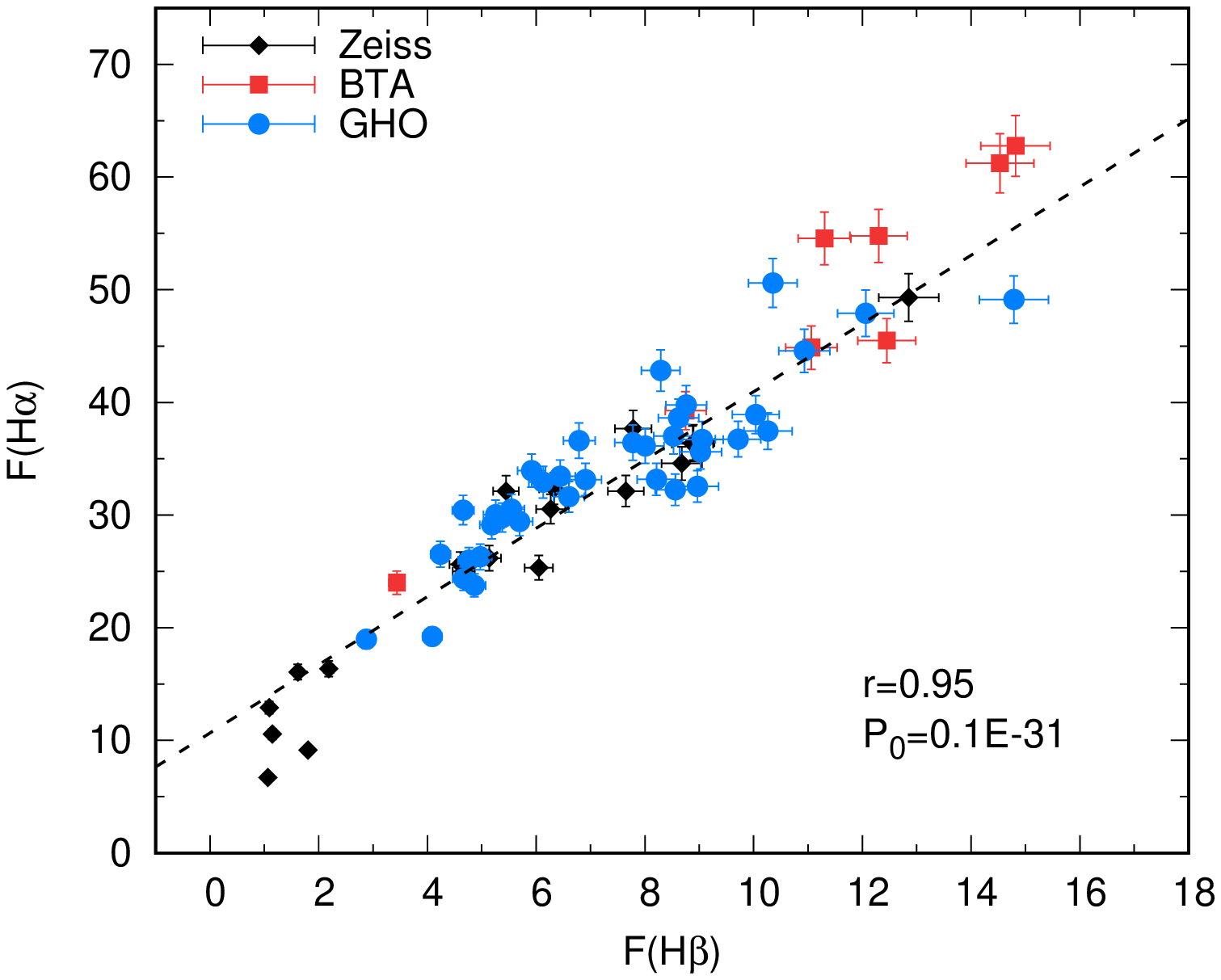}
\includegraphics[width=\columnwidth]{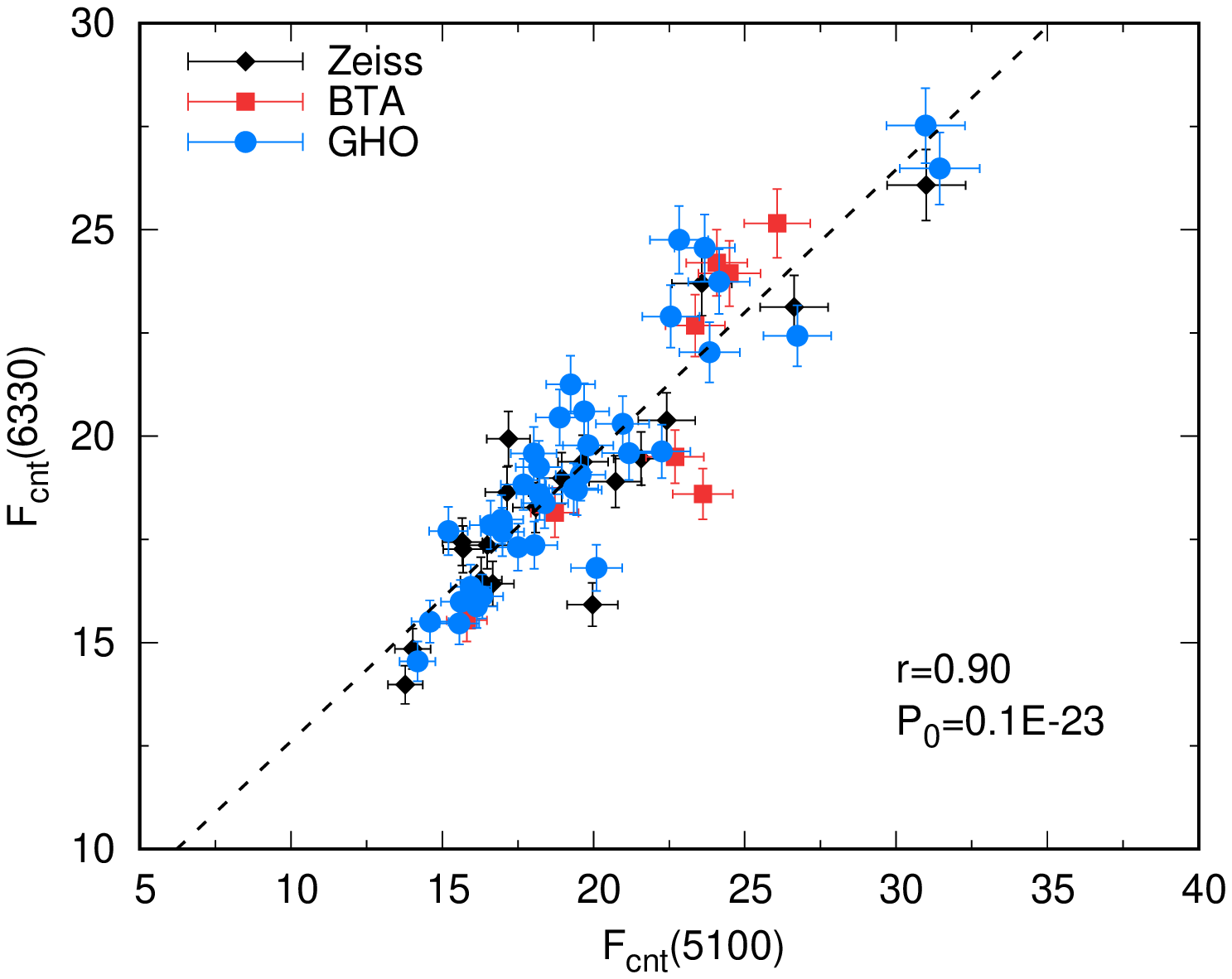}
\caption{H$\alpha$ vs. H$\beta$ line flux (left) and red vs. blue continuum flux (right).
Symbols and units are the same as in Figure \ref{fig2}. The correlation coefficients $r$ and the corresponding null-hypotesis $P_0$ values are also given.} \label{fig4}
\end{figure*}

\subsection{Measurements of the spectral fluxes, their unification and errors}
\label{sec:flux}

Using the  scaled  spectra, we determined the fluxes in the blue and red continuum and in the  broad emission lines for each  data set
(i.e. data with a given aperture and telescope from Table \ref{tab2}).  
The average flux in the continuum near the H$\beta$ line at the observed wavelength 5145 \AA\
($\sim$5100 \AA\ in the rest frame) was obtained  by averaging flux in the spectral range of  (5130--5160) \AA. The continuum near the H$\alpha$ 
line at the observed wavelength  6385 \AA\ ($\sim$6330 \AA\ in the rest frame),  was measured by averaging fluxes in the spectral range of 
(6370--6400) \AA. To measure the observed fluxes of 
H$\alpha$ and H$\beta$, it is necessary to subtract the underlying continuum. 
For this goal, a linear continuum  was fitted through the windows of
20 \AA, located at observed  wavelength 4760 \AA\ and  5120 \AA\ for H$\beta$, and at 6390 \AA\
and 6820 \AA\ for H$\alpha$. After the continuum subtraction, we defined the line fluxes in the following 
observed wavelength bands: from 4845 \AA\ to 4965 \AA\ for H$\beta$ and from 6490 \AA\ to 6750 \AA\ for H$\alpha$.  

In order to investigate the long-term spectral variability of an AGN, it is necessary to gather a consistent and uniformed data set. 
Since observations were carried out using instruments of different apertures, it is necessary to correct the line and continuum 
fluxes for these effects \citep[][]{pe83}. As  reported in our previous papers 
\citep[][]{sh01,sh04,sh08,sh10,sh12,sh13,sh16,sh17}
we determined a point-source correction factor ($\varphi$) and an aperture-dependent correction factor to account for the host galaxy contribution to the continuum (G(g)). We used the following expressions \citep[see][]{pe95}
\begin{eqnarray}
 F({\rm line})_{\rm true} = \varphi \times F({\rm line})_{\rm obs} \\ 
 F({\rm cont})_{\rm true} = \varphi \times F({\rm cont})_{\rm obs} - G(g)
\end{eqnarray}
where index "obs" denotes the observed flux, and "true" the aperture corrected flux. The spectra of the 2.1 m telescope at GHO
(INAOE, Mexico) within an aperture of 2.5\arcsec$\times$6.0\arcsec were adopted as standard (i.e. $\varphi$= 1.0, G(g)=0 by definition). 
The correction factors $\varphi$ and G(g) are determined empirically by comparing pairs of simultaneous observations from each of given
telescope data sets (see Table \ref{tab4}) to that of the
standard data set \citep[as it was used in AGN Watch, see e.g.][]{pe94,pe98,pe02}.
The time intervals between observations 
which were defined as  quasi-simultaneous are typically of 1-3 days. 
 In Table \ref{tab5}, the fluxes for the continuum at the rest-frame wavelengths at 5100 \AA \,
and 6330 \AA, as well as the  H$\beta$ and H$\alpha$ lines and their errors are given. 
The mean errors  of the continuum and line fluxes are in the interval between  3.3\% and 4.5\%. The error-bars were estimated by comparing measured fluxes from the spectra obtained within the time interval that is shorter than 3 days. The flux errors  listed in 
Table \ref{tab5} were estimated using the mean errors.

\begin{table*}
\centering
\caption{The measured continuum and line fluxes, and their estimated errors.
Columns are: (1): Number of spectra, (2): UT-date, (3): Modified Julian Date (MJD), 
(4): Blue continuum, (5): H$\beta$, (6): Red continuum, and (7): H$\alpha$. The line fluxes are in units of $10^{-13} \rm erg \, cm^{-2} s^{-1}$, and continuum fluxes in units of $10^{-15} \rm erg \, cm^{-2} s^{-1}$\AA$^{-1}$. 
The full table is available online as Supporting information.}
\label{tab5}
\begin{tabular}{ccccccc}
\hline
N & UT-date & MJD & F${\rm 5100}\pm \sigma$ & F(H$\beta$)$\pm \sigma$ & F${\rm 6330}\pm \sigma$ & F(H$\alpha$)$\pm \sigma$  \\
1& 2 & 3 & 4&5& 6 & 7  \\
\hline
  1 & 14.01.1996 & 50096.63 & 26.17$\pm$1.10 &  14.41$\pm$0.62 &   -             &   -            \\       
  2 & 20.01.1996 & 50103.47 &      -                    &  -	                          & 26.71$\pm$0.88  & 64.82$\pm$2.92 \\      
  3 & 19.03.1996 & 50162.31 & 23.36$\pm$0.98 &  14.82$\pm$0.64 & 22.68$\pm$0.75  & 62.77$\pm$2.82 \\     
  4 & 20.03.1996 & 50163.35 & 26.07$\pm$1.10 &  14.53$\pm$0.62 & 25.15$\pm$0.83  & 61.22$\pm$2.75 \\     
  5 & 05.10.1997 & 50726.63 & 24.49$\pm$1.03 &  12.30$\pm$0.53 & 23.94$\pm$0.79  & 54.77$\pm$2.46 \\     
  6 & 07.10.1997 & 50728.64 & 24.07$\pm$1.01 &  11.31$\pm$0.49 & 24.20$\pm$0.80  & 54.55$\pm$2.45 \\     
  7 & 20.01.1998 & 50834.34 & 23.62$\pm$0.99 &  11.06$\pm$0.48 & 18.60$\pm$0.61  & 44.87$\pm$2.02 \\     
  8 & 28.01.1998 & 50842.44 & 22.69$\pm$0.95 &  12.45$\pm$0.54 & 19.50$\pm$0.64  & 45.49$\pm$2.05 \\     
  9 & 22.02.1998 & 50867.31 & 20.85$\pm$0.88 &  10.51$\pm$0.45 &      -          &  -             \\      
 10 & 07.05.1998 & 50940.53 & 20.92$\pm$0.88 &  10.09$\pm$0.43 &      -          &  -             \\     
\hline
\end{tabular}
\end{table*}


\section{Data analysis and results}
\label{sec:data}

In this section we present our results. First we shortly give an analysis of the photometric observations, and then of the spectral observations, which contain the continuum and broad line variations.

\subsection{Photometric results}
\label{sec:photores}

Our photometric results are presented in Figure \ref{fig1}, where we show the observations in B, V, and R filters. As one can see in Figure \ref{fig1}, 
the photometric observations show the same variability in all three considered filters.  Since there is a lack of data between 2008 and 2012, we cannot be sure that the maximum in the  light curve was in 2007, but it seems it was close to the maximum. The minimum was in 2014, and also in the following years (2014--2018), there were no  large changes in the photometric data.

The color B-V and V-R diagrams (Fig. \ref{fig1}, two bottom panels)  show that in the high activity phase (2002--2008), the slope of the spectra from  blue to red band  was much  steeper (bluer) than in the minimum phase (2012--2018), when the continuum was almost flat. It is also evident from Figure \ref{fig1}  that for every increase in the brightness (flux) the colors  decreased (i.e. became bluer),  which is expected in case of AGN.

\begin{table}
\centering
\caption{Parameters of the continuum and line variations. Columns are: (1): Analyzed  spectral feature, (2): Total number of spectra, (3): Mean flux, (4): Standard deviation, (5): Ratio of the maximal to minimal flux, (6): Variation amplitude (see text).
Continuum flux is in units of $10^{-15} \rm erg \, cm^{-2} s^{-1}$\AA$^{-1}$
and line flux in units of $10^{-13} \rm erg \ cm^{-2}s^{-1}$. }
\label{tab7}
\begin{tabular}{lccccc}
\hline
Feature & N  & $F$(mean) &  $\sigma$($F$) & $R$(max/min)& $F$(var)\\
1 & 2 & 3 & 4 & 5 & 6  \\
\hline
 cont 6330         &   89 &  19.3   & 3.1   &  2.0   &  0.158 \\
 cont 5100         &  122 &  19.0   & 4.1   &  2.7   &  0.167 \\
 H$\alpha$ - total &   89 &  33.5   & 11.1  &  9.7   &  0.331 \\
 H$\beta$ - total  &  122 &   7.0   & 3.1   & 13.9   &  0.442 \\
\hline
\end{tabular}
\end{table}


\begin{figure*}
\centering
\includegraphics[width=5.6cm]{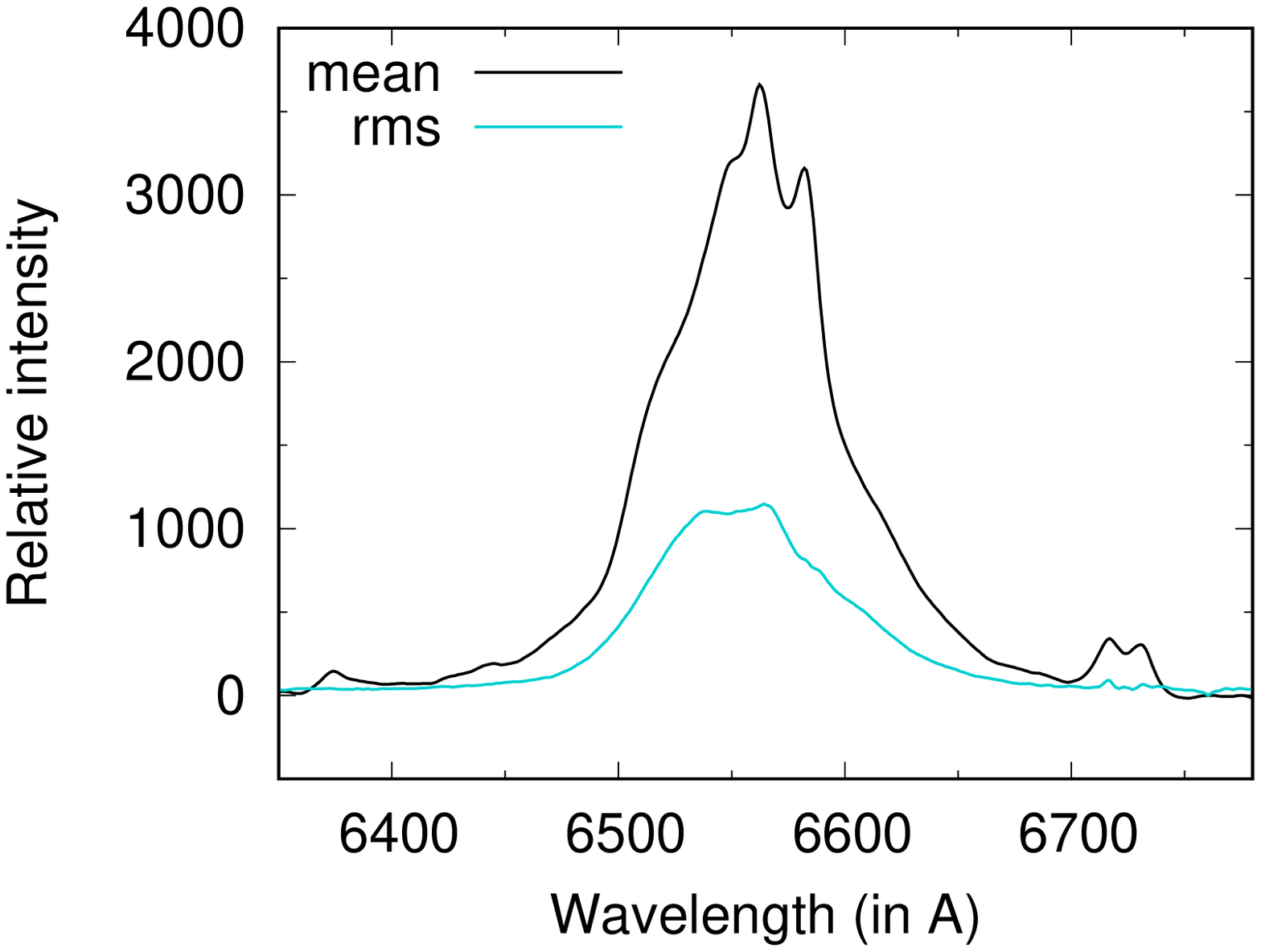}
\includegraphics[width=5.6cm]{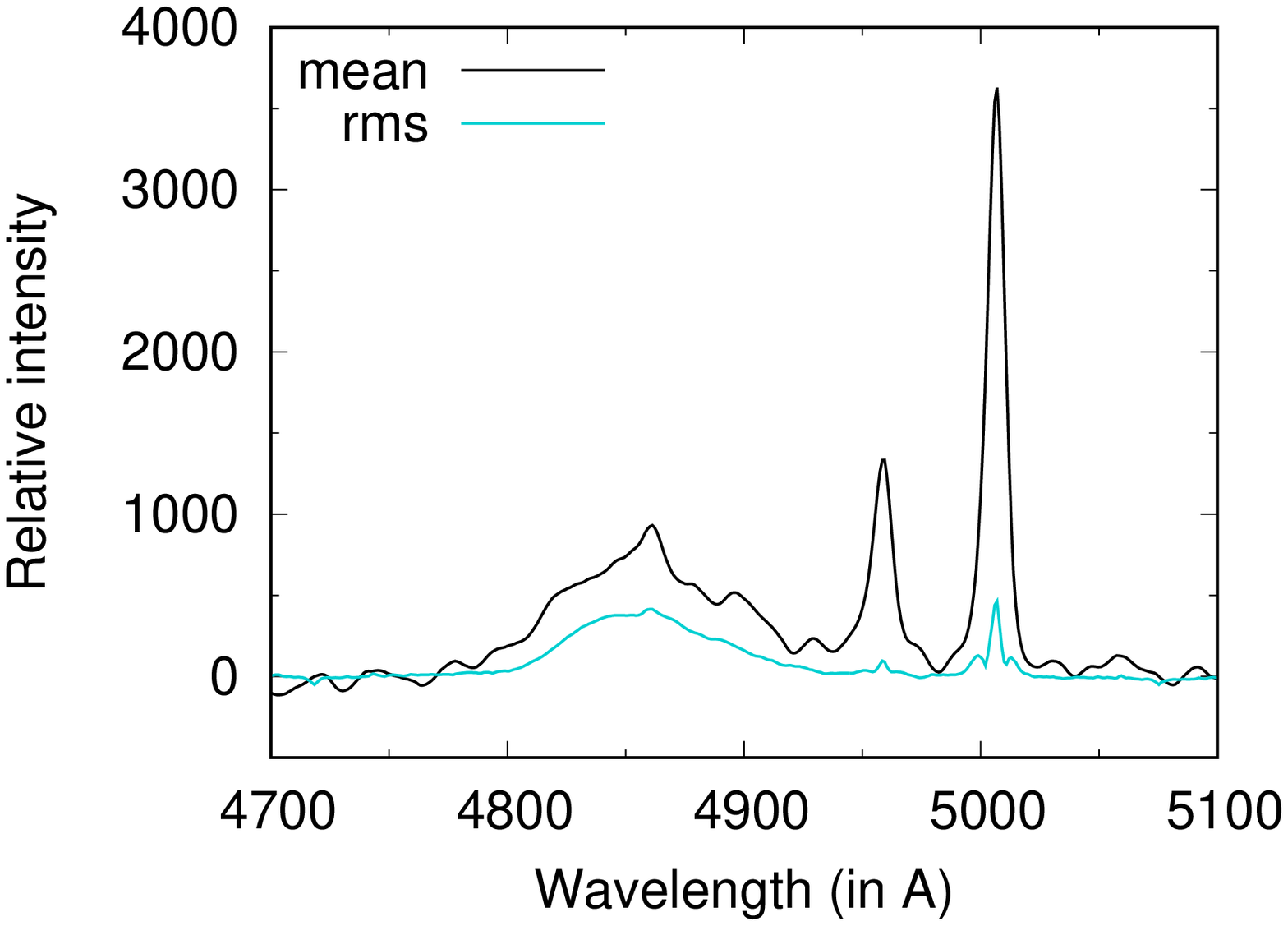}
\includegraphics[width=5.6cm]{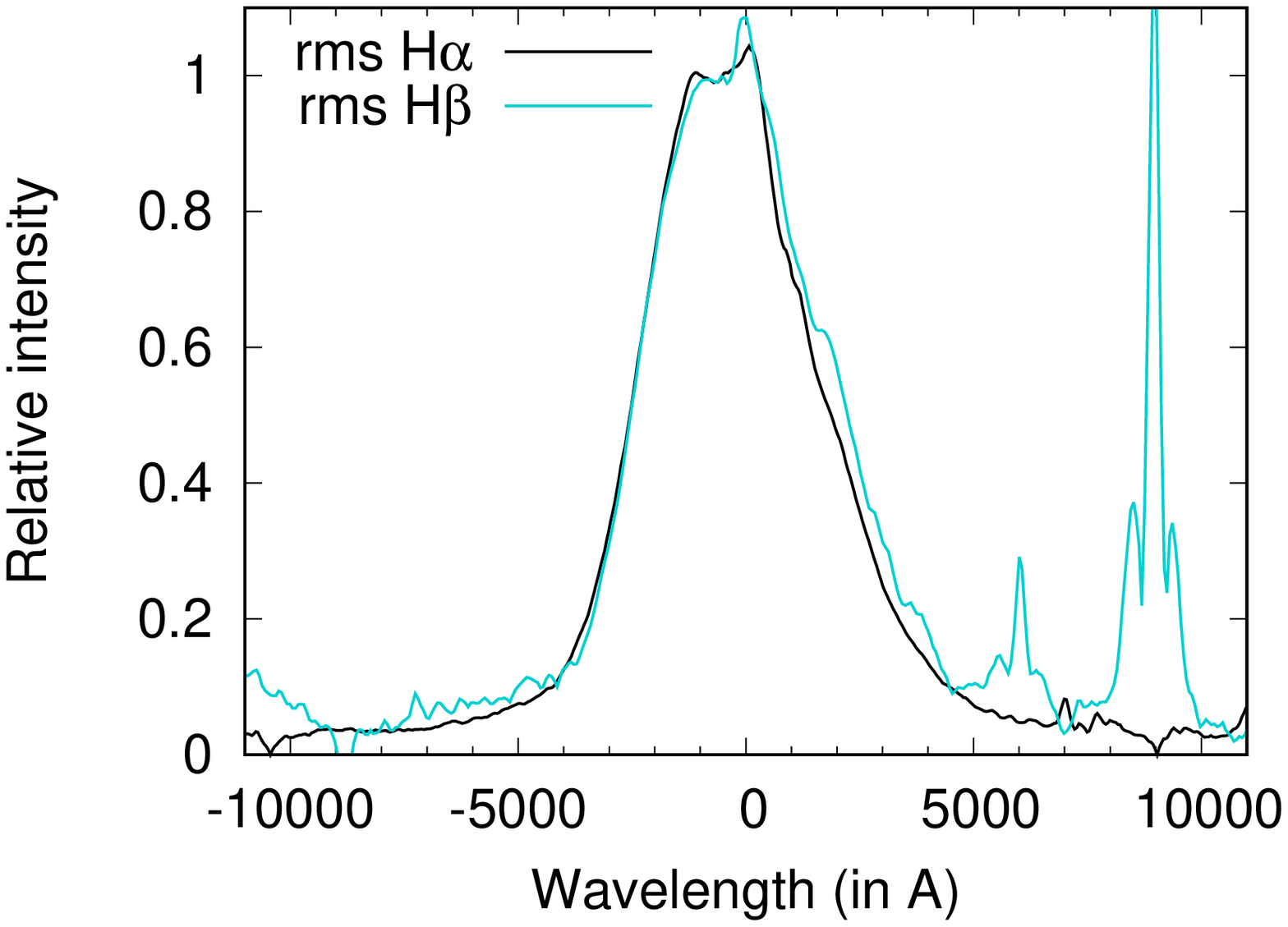}
\caption{Mean and rms-profiles of  H$\alpha$ (left) and H$\beta$ (middle) for spectra
with higher  spectral resolution ($\sim$8\AA). The right panel  represents comparison of the normalized H$\beta$ and H$\alpha$  rms-profiles.} \label{mean}
\end{figure*}

\subsection{Spectral results}
\label{sec:specres}

First we  inspect the spectra obtained during the whole long-term period, finding that the maximum in the optical spectra 
was in 2007, and minimum (as also in photometric observations) in 2014. As it is shown in Figure \ref{fig-mm} we explore the optical spectra
in these two extreme epochs, finding that in the maximum, the continuum is strong,
and broad lines are very prominent, showing 
typical Sy 1 spectrum. In the maximum phase there are Balmer lines
from H$\alpha$ to H$\delta$, and also very  intense Fe II lines, especially,
the Fe II feature between H$\beta$ and H$\gamma$ lines (Fig. \ref{fig-mm}). 

On the other hand, in the minimum phase (Fig. \ref{fig-mm}), the broad lines  disappeared, and the spectrum of NGC 3516 is similar to  Sy 2 spectrum, without strong continuum and broad lines. Additionally, it is interesting that in contrast to a typical Sy 2 spectrum, in the composite spectrum during the minimum, there is no narrow H$\beta$ line, which is probably absorbed. The absorption lines from the host galaxy are dominant, showing both  forbidden and permitted narrow emission lines.

Since we  find an extreme difference between the NGC 3516 spectrum in the phase of minimum and maximum activity, we explore how much the result can be affected by  some artificial effects (as e.g. slit motion). Therefore, we repeated long-slit observations with 6 m telescope of SAO RAS in Febuary 01, 2017 (Fig. \ref{sp2017}), and we find that after subtracting the host galaxy contribution there is a very weak
 H$\beta$ broad component, which cannot be seen in the composite spectra from our monitoring  campaign. 

The light curves of broad line and corresponding continuum fluxes are shown in Figure \ref{fig2},  from which it can be seen that the active phase was more-or-less present in the whole monitored period, beside several last years. In Figure \ref{fig2}, we also plot the  observed fluxes of H$\beta$ and continuum at $\lambda$5100\AA\ reported in  \cite{de18}, which cover only a small part of the period uncovered  by our monitoring campaign.  As it can be seen in Figure \ref{fig2}, the observations of  \citet{de18} fit very well  our photometric and spectral observations.

 In general it is expected that the line flux variation is well correlated with the continuum flux variation, however in some well-known AGNs this is not the case, e.g. in NGC 4151 (see Shapovalova et al. 2008) or Arp 102B (see Shapovalova et al. 2013).  In addition, as it was noted above, NGC 3516  contains a low-luminosity AGN, and it is interesting to explore the response of the line flux to the continuum  flux variability.  To test this, we plot in Figure \ref{fig3} the flux of H$\beta$ (left panel) and H$\alpha$ (right panel) as a function of the continuum flux at 5100 \AA\ and 6330 \AA , respectively. Figure \ref{fig3}  shows that there are good correlations between the line and corresponding continuum (r=0.81 for H$\beta$, and r=0.79 for H$\alpha$), however there is a large scatter especially in the case of the weak broad line flux.  This is expected since in the low activity phase  the weakness of broad emission lines is due to the lack of an ionizing continuum from the nucleus \citep[][]{ki18}.

On the other hand, a better correlation is obtained between the broad H$\alpha$ and H$\beta$ lines ($r=0.95$) and between the blue and red 
continuum ($r=0.90$), which is shown in Figure \ref{fig4}. This is expected, and it confirms that the relative flux calibration of the blue and red spectra obtained from different telescopes was done correctly.

\begin{figure}
\centering
\includegraphics[width=\columnwidth]{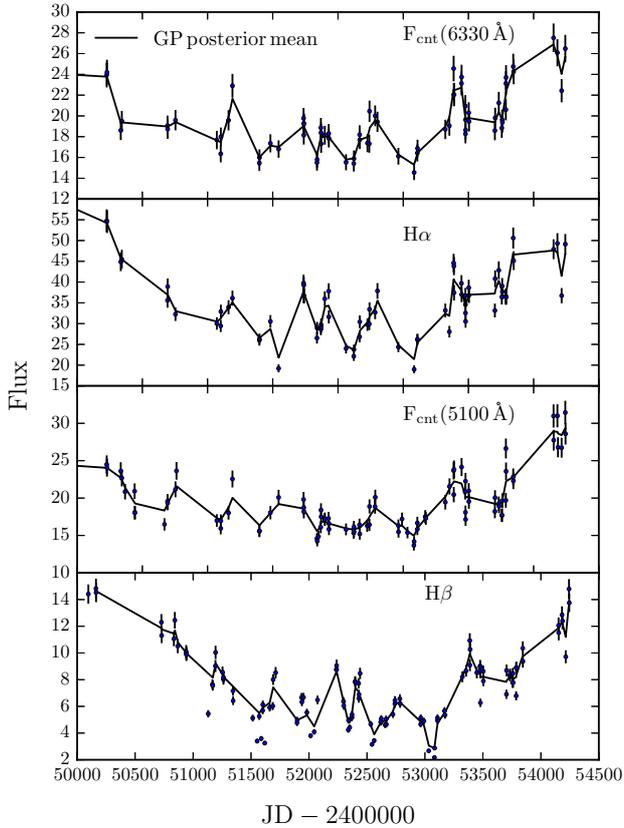}
\caption{GP model fit (solid line) to the observed light curves (points with error bars),
 which are denoted in each plot.  The continuum flux is in units of 
$10^{-15} \rm erg \, cm^{-2} s^{-1}$\AA$^{-1}$ and the line flux in units of $10^{-13} \rm erg \, cm^{-2} s^{-1}$.} \label{gp}
\end{figure}

\begin{figure}
\centering
\includegraphics[width=\columnwidth]{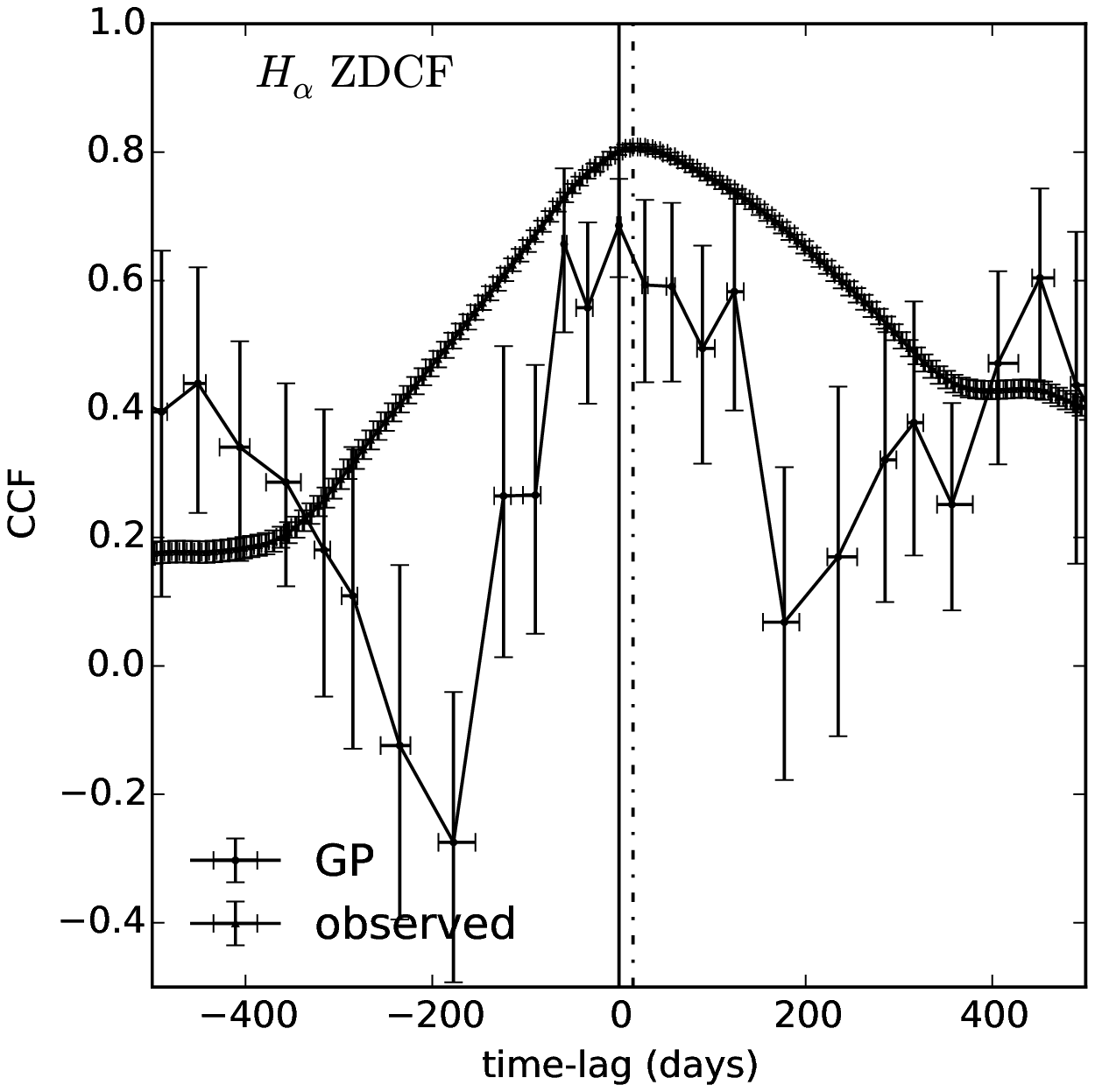}
\includegraphics[width=\columnwidth]{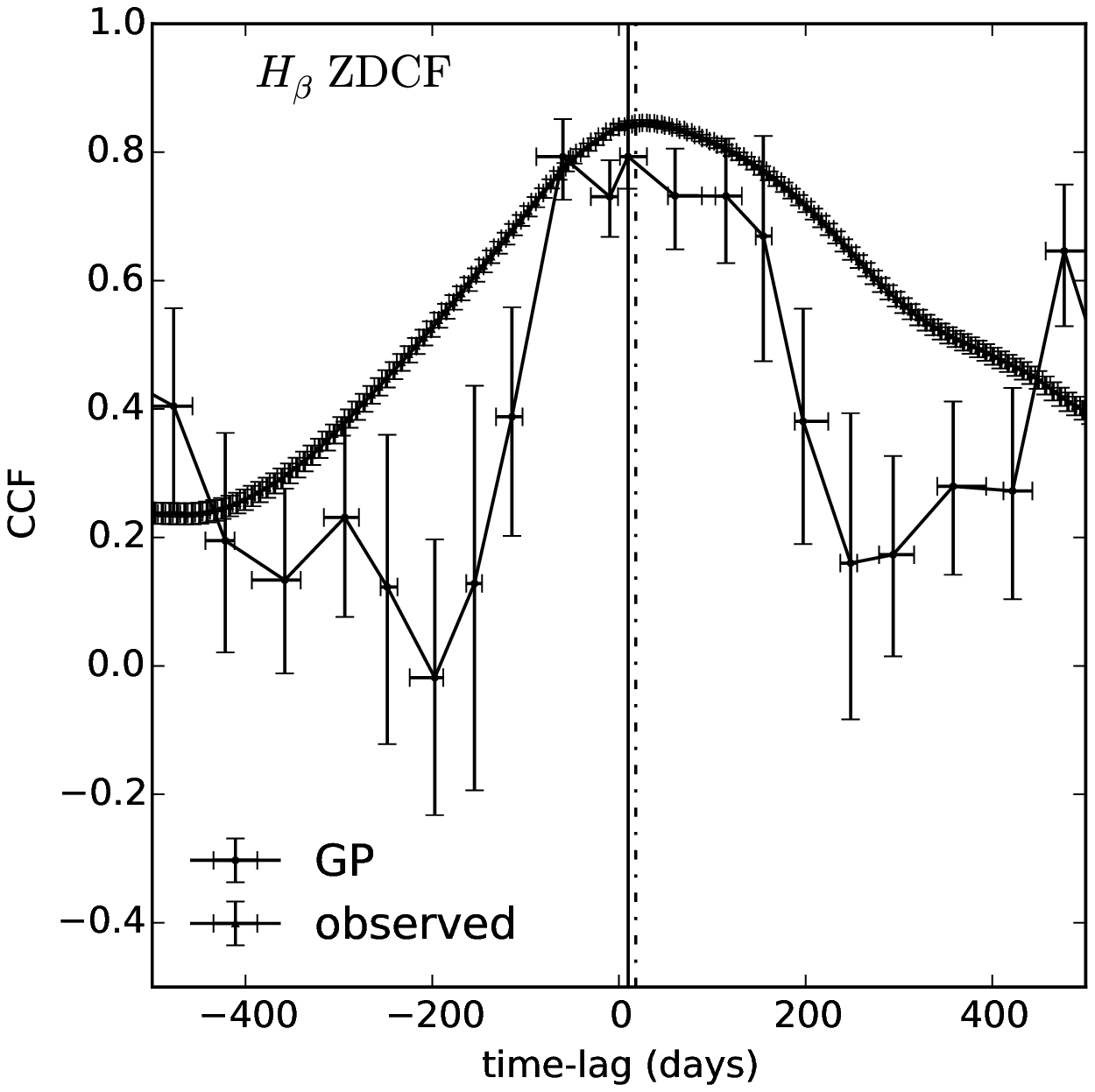}
\caption{Cross correlation functions (ZDCF) for the H$\alpha$ (top) and H$\beta$ (bottom). The error bars  show the  ZDCF  for observed and GP modeled curve.  The vertical lines mark the obtained time lag for the observed (dashed–dotted) and GP modeled light curve (solid).} \label{ccf}
\end{figure}

\subsubsection{Variability of the emission lines and the optical continuum}
\label{sec:var}

As it can be seen in Figures \ref{fig1} and \ref{fig2}, there is a large  variability in the spectra during the  monitored period. 
To explore the rate of variability we calculate the variation amplitude using the method given by \citet{ob98} and present it in Table \ref{tab7}. 
The changes in the continuum were around two times (2.7 times for $\lambda$5100 \AA\ and 2 times for
$\lambda$6330 \AA),  which is usual for Sy 1 galaxies \citep[see e.g.][]{sh17}. However, the line flux  changed for somewhat more than 10 times (Tab. \ref{tab7}), which is expected in AGN which change their type, as e.g. in Fairall 9 in which broad line fluxes changed by more than an order of magnitude also changed its type from Sy 1 to Sy 1.9 \citep[][]{ko85}.

Additionally, there is a big change in the line profiles. We show the mean and  rms-profiles of the broad H$\alpha$ and H$\beta$  lines in Figure \ref{mean},  from which it can be seen that the mean profile of H$\alpha$ and H$\beta$ lines show structures in the blue and red wing, like shoulders which may indicate a complex BLR \citep[see][]{po02}. We  construct the mean and rms-profiles for both lines using only spectra with resolution of 8 \AA\, and find that the full width at half maximum (FWHM) of the mean H$\alpha$ is 3560 km s$^{-1}$ (the FWHM of H$\alpha$ rms-profile is 4110 km s$^{-1}$),  whereas the mean H$\beta$ seems to be broader with the FWHM of 5120 km s$^{-1}$ (the FWHM of H$\beta$ rms-profile is  4360$\pm$80 km s$^{-1}$).  For the FWHM of H$\beta$ rms-profile, which is later used for the black hole mass estimation, we estimated the  uncertainty by making several  measurements for different  levels of the underlying continuum, taking the resulting average for the FHWM and the uncertainty to be  1$\sigma$. Both  mean-profiles and their rms show a red asymmetry, that  may be caused by the inflow or gravitation redshift \citep[see][]{jo16},  but also other effects can be present, as e.g. it could imply outflow if the inward facing side of the BLR clouds are brighter than the outward face, as it is suggested by photoionization modeling. Both rms-profiles have the same shape (see Figure \ref{mean}, right panel),  which indicates similar kinematics of both regions.

\subsection{Time-lag and periodicity analysis}
\label{sec:lag}

The time-lags between light curves in the H$\alpha$, H$\beta$ lines and  corresponding continuum bands are determined from the z-transformed discrete correlation function (ZDCF) analysis \citep[following the technique detailed in][]{al97,po14,sh16,sh17}.
Note that the light curves are sampled at the same time,  as noted in \cite{ed88}
 
 Our long-term observations are covering 22 years, however since there is a large gap after year 2007, in this analysis we used 
 only the part of the light curve up to year 2007 (MJD 54500). In  addition, we modeled Gaussian Process (GP) simulated light-curves, which
are shown in Figure \ref{gp}, in order to obtain the time lags in case of light curves with increased sampling rate. The ZDCF analysis  applied to both  observed and GP simulated light curves, and their ZDCFs  are presented in Figure \ref{ccf}.
Time-lags with the corresponding ZDCFs are given in Table \ref{tab8}. We find the time-lag  of observed H$\alpha$
and continuum $\tau_{\rm zdcf}=0.0^{+2.0}_{-2.0}$ days and  cross correlation coefficient  of $r_{\rm zdcf}=0.69^{+0.07}_{-0.08}$.
Their GP counterparts exhibit larger lag of   $\tau_{\rm zdcf}=15.0^{+5.0}_{-0.0}$ days and similarly larger values
of  $r_{\rm zdcf}=0.81^{+0.01}_{-0.01}$. In the case of H$\beta$ and its continuum the time-lag for observed
light curves is $\tau_{\rm zdcf}=9.7^{+20.3}_{-8.7}$ days, which  $r_{\rm zdcf}=0.79^{+0.05}_{-0.05}$ is slightly
larger then in the case of observed H$\alpha$ (Fig. \ref{ccf}). Their GP counterparts
show the largest time-lag  $\tau_{\rm zdcf}=17.0^{+5.0}_{-0.0}$ days and $r_{\rm zdcf}=0.85^{+0.01}_{-0.01}$.
Results based on the GP light curve analysis  suggest that the time-lag of H$\beta$
could be larger than H$\alpha$ with the upper limit of  about 20 days.

 In addition, the time lags were also calculated with the modified versions of the Interpolated Cross-Correlation Function \citep[ICCF][]{ga86}, as well as the Discrete Cross-Correlation Function \citep[DCCF][]{ed88},  as explained in \cite{pa13}. Both methods produced almost the same time lags for both H$\alpha$ and H$\beta$ lines.

Finally, we generated two artificial  light curves of the duration of 4920 days,  starting from the power spectral density function, with the 30-days cadence and 15 days time lag between them, added the red noise, and  applied the ZDCF method. The ZDCF was able to detect the 15 days time lag with small uncertainty in both cases, with and without red noise. If we randomly extract 70 points from the artificial light curves with red noise, and apply the ZDCF, the obtained time lag is again the same, i.e.  $\tau_{\rm zdcf}=12.0^{+3.1}_{-27.1}$ ($r_{\rm zdcf}=0.96^{+0.01}_{-0.01}$). Therefore we concluded that the sampling rate is influencing more the uncertainty than the estimated time lags. We show that all methods and tests give similar results for the time lags, and later in the text we will use the time lag obtained from the ZDCF method applied on the  GP modeled light-curves  for the calculation  of the mass of the SMBH.

 We note that in the case of long term light-curves, the “red noise problem” could affect the estimated time lags, in such a way that  in addition to the variations on the reverberation timescale, there are longer term variations that bias the estimated lag to larger values, as it was noted by Peterson et al. (2002) for NGC 5548, which is a binary black hole candidate \citep{bo16,li16}. In some cases, the problem of time lag estimates from the "red noise" light curves has been mitigated with the Gaussian process regression \citep{ma10,pa11,te13}, which we also applied here. However, we cannot neglect a possibility that true time lag may be somewhat smaller then our analysis indicates.

\begin{table}
\centering
\caption{The results of the ZDCF analysis. Columns are: (1): Analyzed light curves. (2): Number of points. (3): Time-lag in days  from the ZDCF. (4): ZDCF correlation coefficient.}
\label{tab8}
\begin{tabular}{lccc}
\hline
Light curves & N  & $\tau_{\rm zdcf}$ [days] &  $r_{\rm zdcf}$   \\
1 & 2 & 3 & 4  \\
\hline
GP cnt vs H$\alpha$      &  3728 &  15.0$_{-0.0}^{+5.0}$    &  0.81$_{-0.01}^{+0.01}$ \\
GP cnt vs H$\beta$        &  3728 &  17.0$_{-0.0}^{+5.0}$    &  0.85$_{-0.01}^{+0.01}$ \\
OBS cnt vs H$\alpha$    &   50    &  0.0$_{-2.0}^{+2.0}$     &  0.69$_{-0.08}^{+0.07}$   \\
OBS cnt vs H$\beta$      &   63   &  9.7$_{-8.7}^{+20.3}$     &  0.79$_{-0.05}^{+0.05}$ \\
\hline
\end{tabular}
\end{table}


\subsubsection{Periodicity}
\label{sec:period}

In order to test for any meaningful signal in the light curves,
we calculate for  observed and GP light curves
Lomb-Scargle periodogram with a bootstrap analysis
to assess its significance as it is described in \cite{sh16,sh17}.
We test whether a
purely red noise model can produce  a periodic variability of
light curves. We obtain random
light curves from the Ornstein--Uhlenbeck (OU) process (red noise)
sampled to a regular
time interval.

\begin{figure}
\centering
\includegraphics[width=\columnwidth]{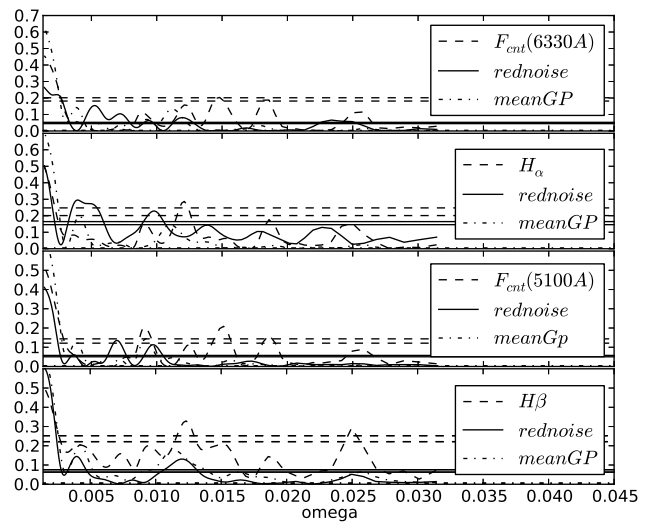}
\caption{Lomb-Scargle periodogram of the observed light curves, red noise and GP
models.
The horizontal lines show the $1\%$ and $5\%$ significance levels for the
highest peak in the periodogram, determined by 1000 bootstrap resamplings.} \label{period}
\end{figure}

The periodogram analysis (Fig. \ref{period}) shows that there are no
significant periodic signals.
The largest peak of each curve corresponds to the whole observed period
 However, in the H$\beta$ line one can see a peak at about $\sim$523  days,  the continuum at $\lambda$5100 \AA\  has a peak at  $\sim$698 days,
whereas  the  H$\alpha$ line  has a peak at $\sim$ 515 days.
On the other hand, GP light curves calculated from the observed light
curves do not exhibit
any significant periodic signal.

\begin{figure*}
\centering
\includegraphics[width=16cm]{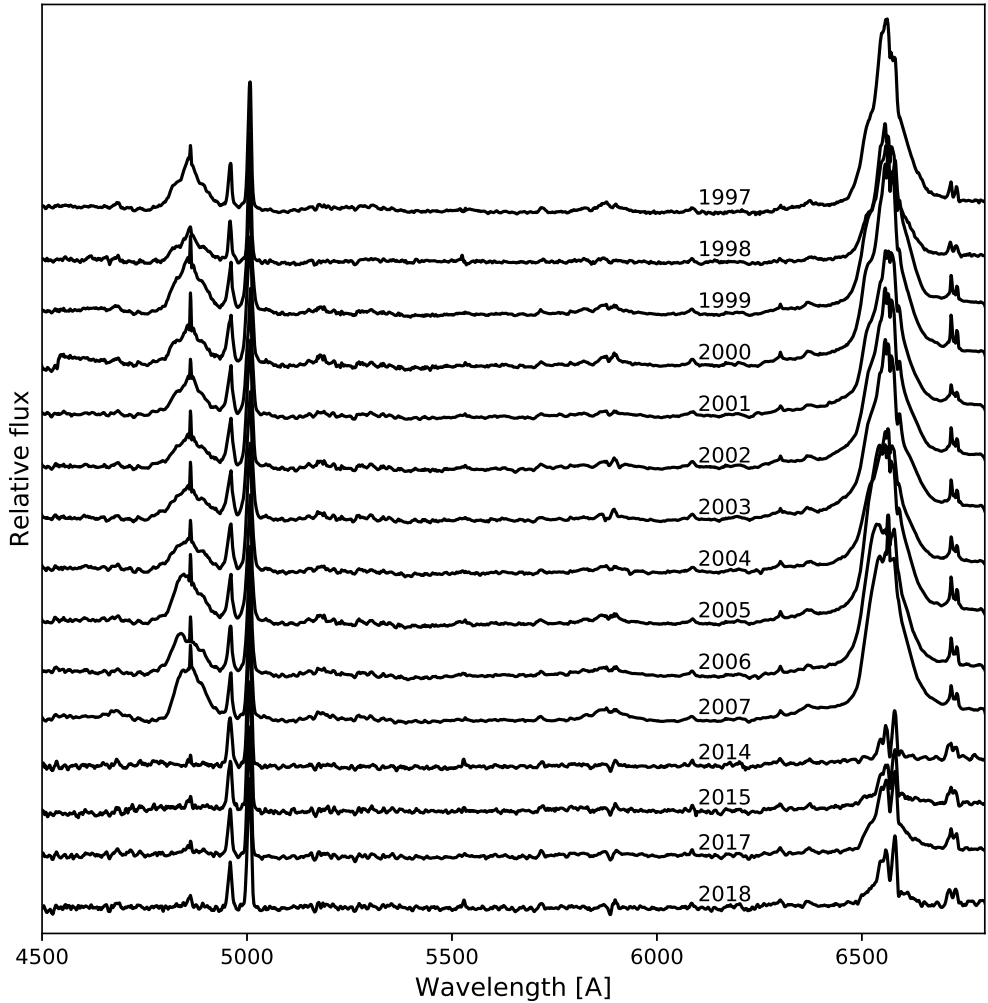}
\caption{Year-average host-galaxy and continuum subtracted spectra, obtained for those spectra covering the total wavelength range. All spectra are normalized to the [O III] $\lambda$5007 intensity and shifted for comparision.} \label{year}
\end{figure*}

\section{Discussion}

In this paper we investigate  the long-term photometric and spectroscopic
variability of  NGC 3516  observed with three telescopes in a monitoring campaign that lasted for 22
years (from 1996 to 2018). We find that the intensity in the broad lines as well as in the continuum
flux  was changing  at a rate between two times in the continuum flux, and more than ten times 
in the broad-line flux. NGC 3516 changed the type of activity during the monitoring campaign, having the typical Sy 1 spectrum in the first period of the campaign, and changing from 2014 to the spectrum which is without broad lines, similar to the spectrum of 
Sy 2 galaxies, which is clearly visible in Figure \ref{year} where we plot the year-average spectra corrected for the host-galaxy and continuum.
We note that, as can be seen in Figure \ref{fig-mm}, the H$\beta$ narrow line also disappeared in the composite spectrum in 2014. It seems that the stellar absorption of H$\beta$ in the low-state phase is too strong that the narrow emission was absorbed, which is clear from the host-galaxy corrected spectrum in which the narrow H$\beta$ is slightly appearing in 2014 (Fig.\ref{year}). Such a low-state was also observed in the X-ray \citep[][]{no16}, since observations in 2013--2014 showed that the X-ray emission in this period was at level of just 5\% of the average flux observed in 1997--2002 period \citep[][]{no16}.

\begin{figure}
\centering
\includegraphics[width=\columnwidth]{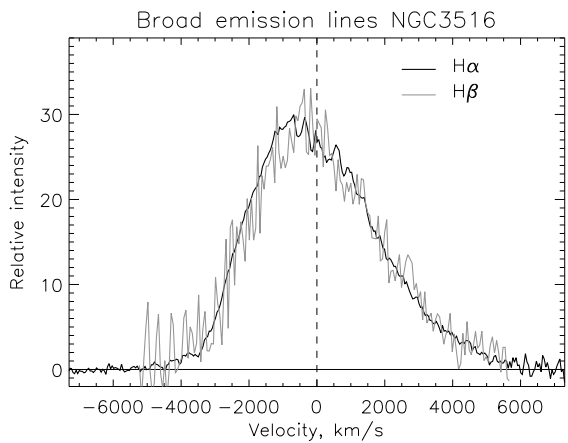}
\caption{The comparison of the broad H$\alpha$ and H$\beta$ profiles observed in
2017  with SCORPIO-2 spectrograph at the 6-m telescope, during the phase near minimal activity.} \label{broad2017}
\end{figure}

As we noted above, we performed observations in 2017 to get the high resolution spectrum of the AGN in the minimum phase (see \S2). Figure \ref{sp2017}  shows that after subtracting the host galaxy spectrum, there are very weak broad emission lines (H$\alpha$, but also H$\beta$). We compare the broad line profiles of H$\beta$ and H$\alpha$ (Fig. \ref{broad2017}) and find that they have practically the same line profiles.  The 
H$\alpha$ and H$\beta$ have also the same FWHM which is around 
2000 km s$^{-1}$ (Fig. \ref{broad2017}). The FWHM from this period is around two times smaller than the FWHM of the averaged line profile, and both broad components are significantly shifted to the blue (around 1000 km s$^{-1}$), which can indicate both the outflow in the minimum phase, and also the disc emission \citep[][]{po02}. The shifted broad H$\beta$ and 
H$\alpha$ with an extensive red wing can be created in an outflowing disc-like BLR
\citep[as it was discussed and presented in Figure 19 of][]{pop11}. The 
investigation of the  broad line profiles and consequently  the model of  the BLR 
structure of NGC 3516 will be given in Paper II.

Additionally, we measure the narrow line ratios in the minimum phase and we obtain
  [O III]$\lambda$5007/H$\beta$=10.3, [N II]$\lambda$6583/H$\alpha$=3.8,  which 
indicates the strong shock wave excitation in  the narrow line region, 
that can be in the process of gas outflowing  on the edge of the 
ionization conus \citep{af07}.

Variability in the line profiles of broad Balmer lines is high, as we noted above, the broad component almost disappeared in the last period of 
the monitoring campaign (in 2014), and after the total minimum,   the low-flux broad lines started to appear, as e.g. in 2017. The line variability in the UV spectral range was reported by
\cite{go99}. They found that high-ionization emission lines 
(Ly$\alpha$ $\lambda$1215, C IV $\lambda$1549, N V $\lambda$1240, and He II $\lambda$1640) showed significant variation that was of the order of a factor of $\sim$2, similar as 
we find in the H$\beta$ and H$\alpha$ lines.

One of the most interesting facts is that NGC 3516 is the changing-look AGN (Fig. \ref{year}), and the nature of these objects can be different
\citep[see][etc.]{ma03,bi05,ki18,no18}. If  an AGN is a changing-look from type 1 to type 1.9 or 2, this can be 
explained via a variable absorption of matter between an observer and
the accretion disc. In that case the obscuring material (e.g. dust clouds) should have a patchy distribution, then the dynamical movement of dust clouds can result in a change of the continuum (and broad line) emission, which affects the current  classification. 

On the other side, any lack of accretion (that may  be caused by different effects) could result in the lack of the ionization continuum, and consequently in the lack of broad emission lines \citep[see][]{ki18,no18}. As e.g.  Mrk 1018 changed from type 1.9 to 1 and returned back to 1.9 over a period of 40 years \citep[][]{ki18}. From our observations we could not see that there was a change in the 
past, since 1996, when we started our monitoring campaign. However, from the
 comparison of the H$\beta$ profile observed in 1943 by \cite{sy43} and 
in 1967 by \citep[][]{and68}, it can be clearly seen that the broad H$\beta$ 
component was present in the observation from 1943, and was absent in the epoch 
of 1967 \citep[see Fig. 3 in][]{and68}. Therefore we can not exclude that 
there is some repetition of the changing look of NGC 3516 (with some periodicity) that should be 
investigated in the future. Additionally,  there is something in common with 
Mrk 1018, since both AGN in the phase of type 1 showed complex broad Balmer lines which indicate more than one emission line region \citep[][]{po02,ki18}.  Finally, we note that NGC 3516 showed a strong absorption in the UV lines
\citep[][]{go99}, as well as in the X-ray continuum \citep[see][etc.]{kr02,tu11,ho12}.

\subsection{Black hole mass determination}

The SMBH mass ($M_{\rm BH}$) of NGC 3516 can be estimated  using the virial theorem \citep[see][]{pe14}:
\begin{equation}
M_{\rm BH}=f{\Delta V_{\rm FWHM}  R_{\rm BLR}\over G},
\end{equation}
where $\Delta V_{\rm FWHM}$ is the   line-of-sight orbital velocity at the radius $R_{\rm BLR}$ of the BLR, which is estimated from the width
 of the variable part of the  H$\beta$ emission line, and $f$ is a factor that depends on the geometry and  orientation of
the BLR. Different values  are obtained for the scale factor $f$,  depending whether it was determined statistically \citep[see e.g.][]{on04,wo15} or by detailed modeling of the reverberation data \citep[see e.g.][]{pa14,gr17}. 

Here we will use the recent result for the $f$ factor from \cite{wo15} who obtained ${\rm log} f = 0.05 \pm0.12$  for the H$\beta$ FWHM-based $M_{\rm BH}$ estimates. Taking into account that the dimension of the H$\beta$ BLR is $\sim$ 17 light days and that the
FWHM of the  H$\beta$  rms-profile is  4360 km s$^{-1}$, we obtained that the central SMBH has a mass of (4.73$\pm$1.40)$\times10^7\ M_\odot$ ($\log(M[M_\odot])=7.67$). The  uncertainty in the time lag and FWHM are propagated through to calculate the formal mass uncertainty.

Although we used the FWHM of the H$\beta$ line in our analysis, we obtain the result which is in agreement with the estimates of other authors who calculated the mass based on the line-dispersion, i.e. \cite{pe04} found that the mass is (4.27$\pm$1.46)$\times 10^7M_\odot$ ($\log(M[M_\odot])=7.63$), \cite{de10} reported the mass of  (3.17$^{+0.28}_{-0.42}$)$\times 10^7M_\odot$($\log(M[M_\odot])=7.50$),  and the most recent finding of \cite{de18} is the mass of  (4.27$\pm$1.35)$\times 10^7M_\odot$ ($\log(M[M_\odot])=7.63$).

\section{Conclusions}

Here we present  the long-term (from 1996 to 2018) photometric and spectroscopic monitoring campaign for NGC 3516.
We analyze observations in order to explore the long-term variability in spectral characteristics of the object. NGC 3516 is known as 
variable object from X-ray \citep[][]{no16} to the optical \citep{de10,de18} spectra. From our analysis of the long-term monitoring we can outline 
following conclusions:

\begin{enumerate}

\item During more than 20 years of monitoring,  the range of continuum flux (blue at 5100 \AA\ and red at 6330 \AA) variations exceeded a factor of two, whereas the range of broad lines variations were  of the order of magnitude.
 This causes a huge  change in the optical spectrum of NGC 3516, that mostly, during the  monitored period has typical Sy 1 spectrum, but from 2014, in the minimum of activity, the broad  lines almost disappeared, and NGC 3516 has the spectrum which is typical  for Sy 2 galaxies, beside the narrow H$\beta$ line, which also disappeared in the composite spectrum after 2014 (Fig.\ref{year}). This indicates a strong absorption in this period. The spectrum did  not change a lot in the following four years, until the end of the monitoring in 2018, when only a weak broad H$\alpha$ and H$\beta$ components are present. These components have the same shape (blue-shifted peaks around 1000 km 
s$^{-1}$ and larger red wing) with smaller FWHM (around 2000 km s$^{-1}$) than averaged 
broad line profiles (FWHM $\sim$ 4000-5000 km s$^{-1}$). This indicates 
that the structure of the BLR was significantly changed.

\item During the main monitoring period (1996–-2007), there is a good correlation between fluxes in the broad lines and  the corresponding continuum ($r\sim0.8$). This indicates that the main mechanism for the formation of broad emission lines in the BLR is the photoionization by the continuum from the nucleus. However, in the low activity phase broad-line fluxes are caused  mainly by shock excitation as a result of an outflow, and not by photoionization from a pure AGN (see  Fig. \ref{fig2} and discussion). 

 \item We find that the BLR has a dimension of 17 light days of H$\beta$. Using this dimension, and the FWHM of H$\beta$  rms-profile we find that the mass of the  central black hole is (4.73$\pm$1.40)$\times 10^7M\odot$, that is in an agreement with previous estimates \citep[][]{de10,de18}.
 
 \item The mean and rms line profile indicate a complex BLR, probably with two components \citep[see][]{po02}, however we did not investigate the broad line profile in more details, and we  leave the investigation of the broad line structure to Paper II.
\end{enumerate}

\section*{Acknowledgements}

The authors thank the anonymous referee for useful comments and suggestions. 
This work was supported by: INTAS (grant N96-0328), RFBR (grants
N97-02-17625 N00-02-16272, N03-02-17123, 06-02-16843, N09-02-01136,
12-02-00857a, 12-02-01237a,N15-02-02101), CONACYT research grants 39560, 54480, 
151494, and 280789 (M{\'e}xico), and the Ministry of Education and Science of Republic of Serbia through the project
Astrophysical Spectroscopy of Extragalactic Objects (176001). 
We especially thank Borisov N.V., Fathulin T., Fioktistova I., Moiseev A., 
Mikhailov V. ,and Vlasyuk V.V. for taking part in the observations. 





\begin{thebibliography}{99}


\bibitem[Adams \& Weedman  (1975)]{ad75} Adams, T. E., and Weedman, D. W. 1975. ApJ 199, 19

\bibitem[Afanasiev et al. (2007)]{af07} Afanasiev, V.~L., Dodonov, S.~N., Khrapov, S.~S., Mustsevoi, V.~V., \& Moiseev, A.~V.\ 2007, Astrophysical Bulletin, 62, 1 
\bibitem[Afanasiev et al. (2018)]{af18} Afanasiev, V.~L., Popovi{\'c}, L.~{\v C}., \& Shapovalova, A.~I.\ 2019, \mnras, 482, 4985
\bibitem[Alexander  (1997)]{al97}
Alexander, T. 1997, ASSL, 218, 163
  \bibitem[Andrillat  (1968)]{and68}
Andrillat, Y.  1968, AJ, 73, 862
\bibitem[Andrillat \& Souffrin (1968)]{an68}
Andrillat Y. \& Souffrin S., 1968, ApJL, 1, 111
 \bibitem[Andrillat  (1971)]{an71}
Andrillat, Y. 1971, ApJL, 8, 161
\bibitem[Begelman  (1985)]{be85}
Begelman, M. C. 1985, in Astrophysics of Active Galaxies and Quasi-Stellar Objects, ed. J.S.Miller (Mill Valley,CA:University Sciences), 411
\bibitem[Bianchi et al. (2005)]{bi05}
Bianchi, S., Guainazzi, M., Matt, G., Chiaberge, M., Iwasawa, K., Fiore, F., Maiolino, R. 2005, A\&A, 442, 185
\bibitem[Blandford \& McKee  (1982)]{bl82}
Blandford, R. D., and McKee, C. F. 1982, ApJ, 225, 419 
\bibitem[Bochkarev et al. (1990)]{bo90}
Bochkarev N.G., Shapovalova A.I., Zhekov S.A. 1990, AJ, 100, 1799
\bibitem[Bochkarev \& Antokhin  (1982)]{bo82}
Bochkarev, N. G., and Antokhin, L.L.  1982, Sov. Astron. Cirk. No. 1228, 5
\bibitem[Boksenberg \& Netzer  (1977)]{bo77}
Boksenberg, A., and Netzer, H. 1977, ApJ, 212, 37
\bibitem[Boksenberg et al.  (1978)]{bo78}
Boksenberg, A., Snijders, M.~A.~J., Wilson, R., et al.\ 1978, \nat, 275, 404 
\bibitem[Bon et al.(2016)]{bo16} Bon, E., Zucker, S., Netzer, H., et al.\ 2016, \apjs, 225, 29
\bibitem[Cherepashchuk \& Lyutyi (1973)]{ch73}
Cherepashchuk, A. M., and Lyutyi, V. M. 1973. ApL, 13, 165
\bibitem[Collin-Souffrin et al.  (1973)]{co73}
Collin-Souffrin, S., Allein, D., and Andrillat, Y. 1973, A\&A, 22, 343
\bibitem[Collin-Souffrin (1980)]{co88}
Collin-Souffrin, S. 1980, In Variability of Stars and Galaxies-Proceedings of the Fifth 
European Regional Meeting in Astronomy, Institute d'Astrophysics, Liege, Belgium, Cl. 11
\bibitem[Connolly et al. (1995)]{co95} 
Connolly, A. J., Szalay, A. S., Bershady, M. A., Kinney, A. L., \& Calzetti, D. 1995, AJ, 110, 1071
\bibitem[Cousins  (1976)]{co76}
Cousins A. W. J. 1976, MNRAS, 81, 25
\bibitem[Crenshaw  (1986)]{cr86}
Crenshaw, D. M. 1986, ApJS, 62, 821
\bibitem[Crenshaw \& Peterson  (1985)]{cr85}
Crenshaw, D. M., and Peterson, B. M. 1985, ApJ, 291, 677 
\bibitem[Denney et al. (2010)]{de10}
	Denney, K. D., Peterson, B. M., Pogge, R. W. et al. 2010, ApJ, 721, 715 
\bibitem[De Rosa et al.  (2018)]{de18}
De Rosa, G., Fausnaugh, M.~M., Grier, C.~J., et al.\ 2018, \apj, 866, 133 
\bibitem[Devereux (2016)]{de16}
Devereux, N., 2016, ApJ, 822, 69
\bibitem[Dietrich et al.  (1998)]{di98}
Dietrich, M., Peterson, B.~M., Albrecht, P., et al.\ 1998, \apjs, 115, 185 
\bibitem[Dimitrijevi\'c et al.  (2007)]{di07}
Dimitrijevi\'c, M. S., Popovi\'c, L. \v C., Kova\v cevi\'c, J., Da\v ci\'c, M., Ili\'c, D.
 2007, MNRAS, 374, 1181
\bibitem[Doroshenko et al.  (2005)]{do05}
Doroshenko V. T., Sergeev S. G., Merkulova N. I., Sergeeva E. A., GolubinskyYu. V., Pronik V. I., Okhmat N. N., 2005, Astrophysics, 48, 304
\bibitem[Holczer \& Behar (2012)]{ho12}
Holczer, T., Behar, E. 2012, ApJ, 747, 71
\bibitem[Edelson, \& Krolik (1988)]{ed88}
   Edelson, R. A., Krolik, J. H. 1988, ApJ, 333, 646
\bibitem[Francis et al.  (1992)]{fr92}
 Francis, P. J., Hewett, P. C., Foltz, C. B., Chaffee, F. H. 1992, ApJ, 389, 476 
\bibitem[Gaskell \& Sparke(1986)]{ga86} Gaskell, C.~M., \& Sparke, L.~S.\ 1986, \apj, 305, 175 
\bibitem[Goad et al.  (1999)]{go99}
Goad, M. R., Koratkar, A. P., Kim-Quijano, J., Korista, K. T., O'Brien, P. T., Axon, D. J. 1999, ApJ, 524, 707
\bibitem[Grier et al. (2017)]{gr17} Grier, C.~J., Pancoast, A., Barth, A.~J., et al.\ 2017, \apj, 849, 146 
\bibitem[Joni\'c et al.  (2016)]{jo16}
Joni\'c, S., Kova\v cevi\'c-Doj\v cinovi\'c, J., Ili\'c, D., Popovi\'c, L. \v C.  2016, Ap\&SS, 361, 101
\bibitem[Kim et al.  (2018)]{ki18} Kim, D.-C.,  Yoon, I., Evans, A. S. 2018, ApJ, 861, 51
\bibitem[Kraemer et al.  (2002)]{kr02} Kraemer, S. B., Crenshaw, D. M., George, I. M., Netzer, H., Turner, T. J., Gabel, J. R. 2002, ApJ, 577, 98
\bibitem[Kollatschny \& Fricke (1985)]{ko85} Kollatschny, W., \& Fricke, K.~J.\ 1985, \aap, 146, L11
\bibitem[Li et al.(2016)]{li16} Li, Y.-R., Wang, J.-M., Ho, L.~C., et al.\ 2016, \apj, 822, 4
\bibitem[Matt et al.  (2003)]{ma03}	
Matt, G., Guainazzi, M., Maiolino, R. 2003, MNRAS, 342, 422
\bibitem[MacLeod et al.(2010)]{ma10} MacLeod, C.~L., Ivezi{\'c}, {\v Z}., Kochanek, C.~S., et al.\ 2010, \apj, 721, 1014
\bibitem[Node \& Done (2018)]{no18} Noda, H., Done, C. 2018, MNRAS.tmp, 1938
\bibitem[Noda et al.  (2016)]{no16}
Noda, H., Minezaki, T., Watanabe, M. et al. 2016, ApJ, 828, 78
\bibitem[O'Brien et al.  (1998)]{ob98}
O'Brien, P.T., Dietrich, M., Leighly, K. et al. 1998, ApJ, 509, 163
\bibitem[Onken et al.  (2004)]{on04}
Onken, C. A., Ferrarese, L., Merritt, D., Peterson, B. M., Pogge, R. W., Vestergaard, M., Wandel, A. 2004, ApJ, 615, 645
\bibitem[Onken et al.  (2003)]{on03}
Onken, C. A., Peterson, B. M., Dietrich, M., Robinson, A., Salamanca, I. M.
2003, ApJ, 585, 121
\bibitem[Osterbrock  (1977)]{os77}
Osterbrock, D. E. 1977, ApJ, 215, 733
\bibitem[Osterbrock \& Ferland (2006)]{of06} 
Osterbrock, D.~E., \& Ferland, G.~J.\ 2006, Astrophysics of gaseous nebulae and active galactic nuclei, 2nd.~ed.~by D.E.~Osterbrock and G.J.~Ferland.~Sausalito, CA: University Science Books, 2006
\bibitem[Pancoast et al.(2011)]{pa11} Pancoast, A., Brewer, B.~J., \& Treu, T.\ 2011, \apj, 730, 139
\bibitem[Pancoast et al. (2014)]{pa14} Pancoast, A., Brewer, B.~J., Treu, T., et al.\ 2014, \mnras, 445, 3073 
\bibitem[Pati{\~n}o-{\'A}lvarez et al. (2013)]{pa13} Pati{\~n}o-{\'A}lvarez, V., Carrami{\~n}ana, A., Carrasco, L., \& Chavushyan, V.\ 2013, Fermi Symposium eConf Proceedings, C121028, arXiv:1303.1898 
\bibitem[Penston et al.  (1971)]{pe71}
Penston M. J., Penston M. V., Sandage A., 1971, PASP, 83, 783
\bibitem[Peterson (1993)]{pe93}
Peterson, B. M. 1993, PASP, 105, 247
\bibitem[Peterson  (2014)]{pe14}
Peterson, B. M. 2014, SSRv, 183, 253
\bibitem[Peterson \& Collins  (1983)]{pe83}
Peterson, B. M., Collins I. I. G.W. 1983, ApJ, 270, 71
\bibitem[Peterson et al.  (1994)]{pe94}
Peterson, B. M., Berlind, P., Bertram, R., et al. 1994, ApJ, 425, 622
\bibitem[Peterson et al. (1995)]{pe95}
Peterson, B. M., Pogge, R. W., Wanders, I., Smith, S. M., \& Romanishin, W. 1995,
PASP, 107, 579
\bibitem[Peterson et al.  (1998)]{pe98}
Peterson, B. M., Wanders, I., Bertram, R., et al. 1998, ApJ, 501, 82
\bibitem[Peterson et al. (2002)]{pe02}
Peterson, B. M., Berlind, P., Bertram, R., et al. 2002, ApJ, 581, 197
\bibitem[Peterson et al. (2004)]{pe04} Peterson, B.~M., Ferrarese, L., Gilbert, K.~M., et al.\ 2004, \apj, 613, 682
\bibitem[Popovi\'c et al.  (2002)]{po02} 
Popovi\'c, L. \v C,, Mediavilla, E. G.., Kubi\v cela, A., Jovanovi\'c, P. 2002, A\&A, 390, 473.
\bibitem[Popovi\'c et al.  (2014)]{po14}
Popovi\'c, L. \v C., Shapovalova, A. I., Ili\'c, D., Burenkov, A. N., Chavushyan, V. H., Kollatschny, W., 2014, A\&A, 572, A66
\bibitem[Popovi\'c et al.  (2011)]{pop11}
Popovi\'c, L. \v C., Shapovalova, A. I., Ili\'c, D., Kova\v cevi\'c, A., 
Kollatschny, W., Burenkov, A. N.,  Chavushyan, V. H.,  Bochkarev, N. G., 
Le\'on-Tavares, J. 2011, A\&A, 528A, 130
\bibitem[Rees  (1984)]{re84}
Rees, M.J. 1984 1984, ARA\&A, 22, 471
\bibitem[Shapovalova et al.  (2001)]{sh01}
Shapovalova, A. I., Burenkov, A. N., Carrasco, L., et al. 2001, A\&A, 376, 775
\bibitem[Shapovalova et al.  (2004)]{sh04}
Shapovalova, A.I., Doroshenko, V.T., Bochkarev, N.G, et al. 2004, A\&A, 422, 925
\bibitem[Shapovalova et al.  (2008)]{sh08}
Shapovalova, A.I., Popovi\'c, L. \v C., Collin, S., et al. 2008, A\&A, 486, 99
\bibitem[Shapovalova et al.  (2010)]{sh10}
Shapovalova, A. I., Popovi\'c, L. \v C., Bochkarev, N.G., et al. 2010, A\&A, 517A, 42
\bibitem[Shapovalova et al.  (2012)]{sh12}
Shapovalova, A.I., Popovi\'c, L. \v C., Burenkov, A. N., et al. 2012, ApJS, 202, 10
\bibitem[Shapovalova et al.  (2013)]{sh13}
Shapovalova, A. I., Popovi\'c, L. \v C., Bochkarev, N.G., et al., 2013, A\&A, 559A, 10S
\bibitem[Shapovalova et al.  (2016)]{sh16}
Shapovalova, A.I., Popovi\'c, L. \v C., Chavushyan, V., et al. 2016, ApJS, 222, 25
\bibitem[Shapovalova et al.  (2017)]{sh17}
Shapovalova, A.I., Popovi\'c L. \v C., Chavushyan V, Afanasiev, V.L., Ili\'c D., Kova\v cevi\'c  A., et al.
2017, MNRAS 466, 4759
  \bibitem[Seyfert  (1968)]{sy43}
	Seyfert, C. K. 1943, ApJ, 97, 28
\bibitem[Souffrin  (1968)]{so68}
 Souffrin, S. 1968, AJ, 73, 897
\bibitem[Sturm et al. (2018)]{st18} Sturm, E., Dexter, J., Pfuhl, O., et al. 2018, Nature, 563, 657 
\bibitem[Tewes et al.(2013)]{te13} Tewes, M., Courbin, F., \& Meylan, G.\ 2013, \aap, 553, A120
\bibitem[Turner et al. (2011)]{tu11} 
Turner, T. J., Miller, L., Kraemer, S. B., Reeves, J. N. 2011, ApJ, 733, 48
\bibitem[Vanden Berk et al. (2006)]{vb06}  
 Vanden Berk, D. E., Shen, J., Yip, C.-W. et al. 2006, AJ, 131, 84
 \bibitem[Vlasyuk  (1993)]{vl93}
Vlasyuk V. V., 1993, Bull. Spec. Astrophys. Obs., 36, 107
\bibitem[Wanders \& Horne  (1994)]{wa94}
Wanders, I., Horne, K. 1994, A\&A, 289, 76
\bibitem[Wanders et al.  (1993)]{wa93}
Wanders I. van Groningen, E., Alloin, D., et al., 1993, A\&A, 269, 39
\bibitem[Woo et al.(2015)]{wo15} Woo, J.-H., Yoon, Y., Park, S., Park, D., \& Kim, S.~C.\ 2015, \apj, 801, 38
\bibitem[Yip et al. (2004a)]{yi04a}  
Yip, C. W., Connolly, A. J., Szalay, A. S. et al. 2004a, AJ, 128, 585
\bibitem[Yip et al. (2004b)]{yi04b}  
Yip, C. W., Connolly, A. J., Vanden Berk, D. E. et al. 2004b, AJ, 128, 2603
 

\end{thebibliography}








\bsp	
\label{lastpage}
\end{document}